\documentclass[sigconf,nonacm]{acmart}
\AtBeginDocument{%
  \providecommand\BibTeX{{%
    \normalfont B\kern-0.5em{\scshape i\kern-0.25em b}\kern-0.8em\TeX}}}

\usepackage{subfiles}
\usepackage{booktabs}
\usepackage{multirow}
\usepackage{stfloats}
\usepackage{lipsum}
\usepackage{arydshln}
\usepackage{graphicx}
\usepackage{subcaption}
\usepackage[normalem]{ulem}

\usepackage{amssymb}
\usepackage{tikz}
\usepackage{amsmath}
\usepackage{enumitem}
\usepackage{xcolor}
\usepackage{soul}

\newcommand{\vic}[1]{\textcolor{black}{#1}}

\begin{document}

\title{Leveraging Intra-modal and Inter-modal Interaction for Multi-Modal Entity Alignment}

% \affiliation{%

% \author{Anonymous Authors}

\author{Zhiwei Hu}
\affiliation{
  \institution{School of Computer and Information Technology \\ Shanxi University}
  \city{Taiyuan}
  \country{China}
}
\email{zhiweihu@whu.edu.cn}

\author{Víctor Gutiérrez-Basulto}
\affiliation{
  \institution{School of Computer Science and Informatics\\ Cardiff University}
  \city{Cardiff}
  \country{UK}
}
\email{gutierrezbasultov@cardiff.ac.uk}

\author{Zhiliang Xiang}
\affiliation{
  \institution{School of Computer Science and Informatics \\ Cardiff University}
  \city{Cardiff}
  \country{UK}
}
\email{xiangz6@cardiff.ac.uk}

\author{Ru Li}
\authornote{Contact Authors.}
\affiliation{
  \institution{School of Computer and Information Technology \\ Shanxi University}
  \city{Taiyuan}
  \country{China}
}
\email{liru@sxu.edu.cn}

\author{Jeff Z. Pan}
\authornotemark[1]
\affiliation{
  \institution{ILCC, School of Informatics\\ University of Edinburgh}
  \city{Edinburgh}
  \country{UK}
}
\email{j.z.pan@ed.ac.uk}

%% You do not have to enter your paper ID


\begin{abstract}

\vic{Multi-modal entity alignment (MMEA) aims to identify equivalent entity pairs across different multi-modal knowledge graphs (MMKGs). Existing approaches focus on how to better encode and aggregate information from different modalities. However, it is not trivial to leverage multi-modal knowledge in entity alignment due to the modal heterogeneity. In this paper, we propose a  \textbf{M}ulti-Grained  \textbf{I}nteraction framework for \textbf{M}ulti-Modal \textbf{E}ntity \textbf{A}lignment (\textbf{MIMEA}), which effectively realizes multi-granular interaction within the same modality or between  different modalities. MIMEA is composed of four modules: i)  a \textit{Multi-modal Knowledge Embedding} module, which extracts modality-specific representations with multiple individual encoders; ii) a \textit{Probability-guided Modal Fusion} module, which employs a probability guided approach to integrate uni-modal representations  into joint-modal embeddings, while considering the interaction between uni-modal representations; iii) an \textit{Optimal Transport Modal Alignment} module, which introduces an optimal transport mechanism to encourage the interaction between uni-modal and joint-modal embeddings; iv) a \textit{Modal-adaptive Contrastive Learning} module, which distinguishes the embeddings of equivalent entities from those of non-equivalent ones, for each modality. Extensive experiments conducted on two real-world datasets demonstrate the strong performance of MIMEA compared to the SoTA.} \vic{Datasets and code have been submitted as supplementary materials.}

\end{abstract}






\maketitle

\section{Introduction}
\begin{figure}[t!]
    \centering
    \includegraphics[width=0.48\textwidth]{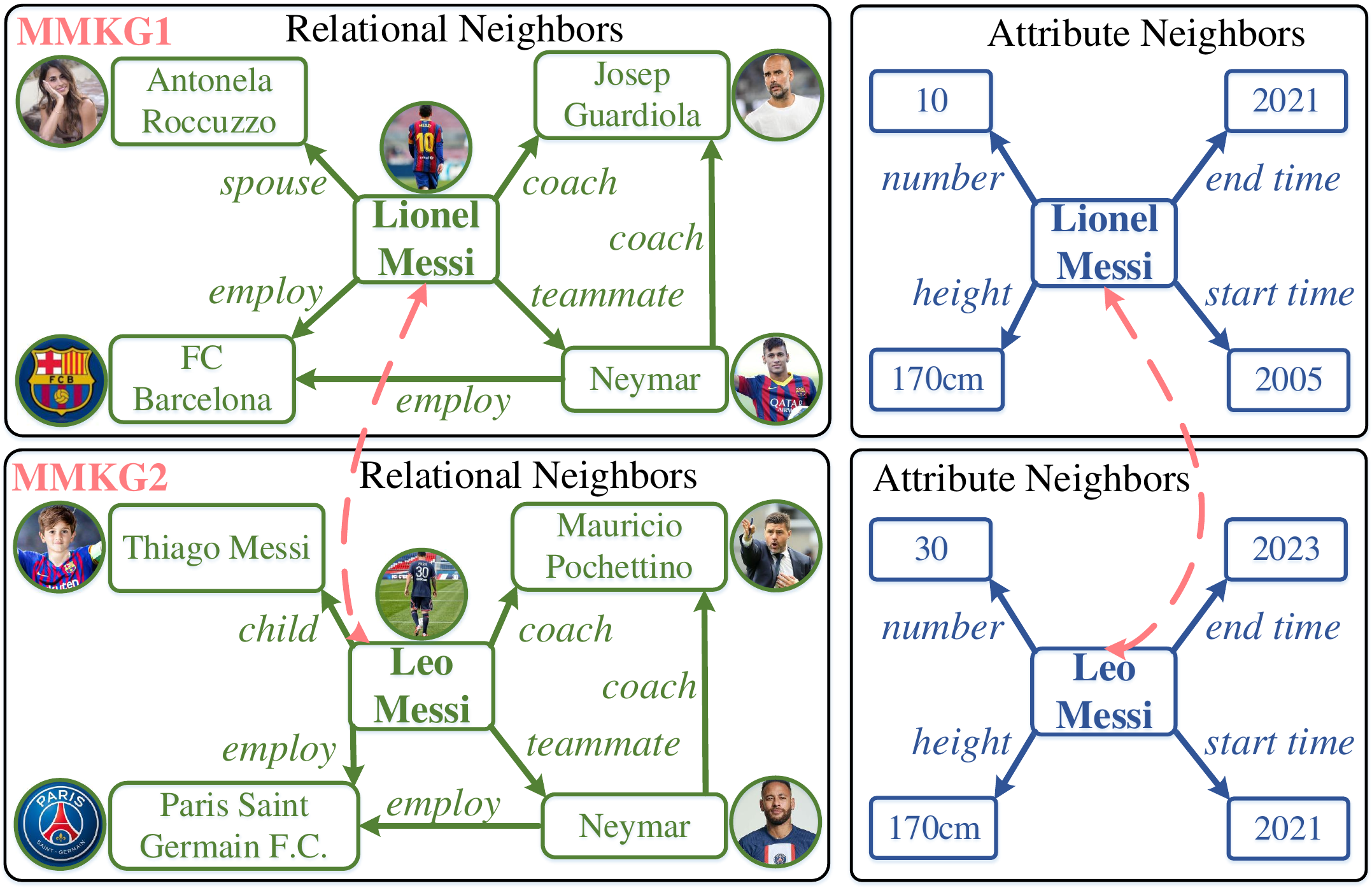}
    \caption{The MMEA task between MMKG1 and MMKG2, aligning the entities \textit{Lionel Messi} and  \textit{Leo Messi}.} %relational neighbors, attribute neighbors, and visual image information should be considered simultaneously.}
    \label{figure_instance}
\end{figure}

\vic{Knowledge graphs (KGs), such as DBpedia~\citep{Jens_2015} and YAGO~\citep{Farzaneh_2015}, employ a graph structure to organize real-world factual knowledge. They provide the backbone of various web-based applications like query answering~\citep{Zhanqiu_2021, Zhiwei_2022, Chau_2023} and search~\citep{NguyenVNNP19,GZCL_2019}. %recommendation systems~\citep{Yuhao_2022, Ding_2022}. 
Recently, several works have extended KGs with additional modeling capabilities, %~\cite{Zhiwei_2023}\nb{V:add references for  KGs+numeric values, KGs+ literals, any other interesting extensions?}, 
as required by different applications. \emph{Multi-modal Knowledge Graphs (MMKGs)}  extend traditional KGs with multi-modal information, e.g.\ visual information. However, like traditional KGs, MMKGs suffer from incompleteness and low coverage. Thus, the  integration of independently developed MMKGs  is  paramount. A key task for MMKG integration is \emph{multi-modal entity alignment (MMEA)}, which aims to identify  equivalent entity pairs  in different MMKGs by taking into account the structure of  MMKGs, as well as the attribute and visual information of entities, see e.g. Fig.~\ref{figure_instance}. 
%
%For example, as shown in Figure~\ref{figure_instance}, to align the entity \textit{Lionel Messi} and the entity \textit{Leo Messi} from different MMKG, knowledge such as relation modal (\textit{Lionel Messi, employ, FC Barcelona}) in MMKG1 and (\textit{Leo Messi, employ, Paris Saint Germain F.C.}) in MMKG2, attribute modal (\textit{Lionel Messi, number, 10}) in MMKG1 and (\textit{Leo Messi, number, 30}) in MMKG2, visual modal an image of Messi wearing the number 10 jersey of Barcelona Club in MMKG1 and an image of Messi wearing the number 30 jersey of Paris Saint Germain Club in MMKG2, need to be considered at the same time.
%
In this way, MMEA facilitates the exchange of knowledge among different MMKGs.}

\vic{A wide variety of approaches to MMEA have been already introduced. Initial proposals~\citep{Ye_2019,Liyi_2020,Fangyu_2021,Hao_2021} 
%MMEA~\citep{Liyi_2020}, EVA~\citep{Fangyu_2021}, and HMEA~\citep{Hao_2021} 
concentrated on the construction of distinct multi-modal fusion modules to integrate entity representations from multiple modalities into joint embeddings and then use aggregated embeddings to predict alignments. A shortcoming of these methods is that they only explore the use of diverse multi-modal representations to enhance the contextual embedding of entities,  overlooking  the capabilities of inter-modal  representations to capture certain types of interactions. 
%In fact, the mutual influence of different modal information plays an important role in multi-modal representation and fusion process. 
To overcome this, some works~\citep{Liyi_2022,Zhuo_2022,Zhenxi_2022} 
%such as MSNEA~\citep{Liyi_2022}, MEAformer~\citep{Zhuo_2022} and MCLEA~\citep{Zhenxi_2022} introduce the 
use siamese networks, transformer mechanisms or  contrastive learning strategies to 
enhance  multi-modal knowledge by exploiting inter-modal interaction.
%comprehensively leverage \textcolor{red}{the multi-modal knowledge by the exploitation of inter-modal effect}\nb{V: I do not understand the red sentence}. 
However, existing frameworks for MMEA still suffer from serious shortcomings:}
%A wide variety of approaches to MMEA have been already proposed, some methods like PoE~\citep{Ye_2019}, MMEA~\citep{Liyi_2020}, EVA~\citep{Fangyu_2021}, and HMEA~\citep{Hao_2021} concentrate on construct distinct multi-modal fusion modules to integrate entity representations from multiple modalities into joint embeddings and then use aggregated embeddings to predict alignments. Although these methods have achieved promising performance, they only explore the use of diverse multi-modal representations to enhance the contextual embedding of entities, serious neglect the interaction capabilities of inter-modal in modeling. In fact, the mutual influence of different modal information plays an important role in multi-modal representation and fusion process. Therefore, some methods such as MSNEA~\citep{Liyi_2022}, MEAformer~\citep{Zhuo_2022} and MCLEA~\citep{Zhenxi_2022} introduce the siamese network, transformer mechanism and contrastive learning strategy to comprehensively leverage the multi-modal knowledge by the exploitation of inter-modal effect. However, it is still not a trivial task due to the following reasons:
\begin{enumerate}[itemsep=0.5ex, leftmargin=5mm]
\item \textbf{Modality Distinctiveness.} 
\vic{Existing methods have difficulties  to explicitly distinguish the importance of each modality. In fact, among all modalities, the  structural modal knowledge is the most prevalent. For instance, the FB15K-DB15K dataset has a total of 714,720 structural triples, while it only has 1,624 relation categories, 341 attribute categories, and 26,281 images. Whether we look at the provided data ratios or the results of  ablation experiments, it is evident that the structural modality provides a richer source of knowledge and therefore deserves more attention. }

%Existing methods are difficult to explicitly distinguish the importance of each modality, i.e., there is no primary or secondary distinction between modalities. In fact, among all modalities, the importance of structure modal knowledge is much higher than that of other modalities. Taking the FB15K-DB15K dataset as an example, it consists of a total of 714,720 triples related to structure modality, only 1,624 relation categories related to relation modality, 341 attribute categories related to attribute modality, and 26,281 images related to visual modality. Whether we look at the provided data quantities or the results of the final ablation experiments, it is evident that structure modality provides a richer source of knowledge and therefore deserves more attention. 
\item \textbf{Modality Interaction Diversity.} \vic{Existing models place more emphasis on the interaction between uni-modal embeddings while overlooking interactions between uni-modal and joint-modal embeddings, leading to a lack of diversity in modality interactions. We advocate that, in practice,  it is necessary to design  mechanisms that better capture the interaction between uni-modal and joint-modal embeddings to fully harness the potential of all available modalities. 
Indeed, the interaction between the joint-modal and uni-modal representations enables simultaneous interactions with more than two modalities, covering  the information  gaps left by only looking at pairwise  interactions.}
% For instance, for the joint-modal embedding incorporating the relational triple  (\textit{Lionel Messi, employ, FC Barcelona}), the attribute triple (\textit{Lionel Messi, number, 10}), and the  image of Messi wearing the Barcelona jersey number 10, existing methods can only perform pairwise uni-modal interactions. 

%For inter-modal interactions, existing models place greater emphasis on the connections between uni-modal while overlooking interactions between uni-modal and joint-modal embeddings, leading to a lack of diversity in modality interactions. In reality, it is necessary to design reasonable interaction mechanism between uni-modal and joint-modal to fully harness the potential of modality. For instance, for the joint-modal incorporating relation modal (\textit{Lionel Messi, employ, FC Barcelona}), attribute modal (\textit{Lionel Messi, number, 10}), and visual modal an image of Messi wearing the number 10 jersey of Barcelona Club, existing methods can only perform pairwise uni-modal interactions. However, the interaction between the joint-modal and uni-modal enables simultaneous interactions with more than two modalities, compensating for the information coverage gaps caused by pairwise modality interactions.
\end{enumerate}
\vic{To address the above two shortcomings, we propose the  method \textbf{MIMEA}, a \textbf{M}ulti-Grained \textbf{I}nteraction framework for \textbf{M}ulti-Modal \textbf{E}ntity \textbf{A}lignment. Specifically, MIMEA includes the following four modules. The \textit{Multi-modal Knowledge Embedding} module utilizes multiple individual encoders to obtain modality-specific representations for each entity. To effectively combine multi-modal knowledge, the \textit{Probability-guided Modal Fusion} module takes structural knowledge as the core, and employs a probability distribution mechanism to integrate uni-modal information into joint-modal representations.  Furthermore, we introduce an \textit{Optimal Transport Modal Alignment} module to capture the interaction between uni-modal and joint-modal embeddings. The integration of the \textit{Probability-guided Modal Fusion} and the \textit{Optimal Transport Modal Alignment} modules realizes inter-modal interactions between uni-modal and joint-modal embeddings. Moreover, we introduce an intra-modal contrastive loss to distinguish the embeddings of equivalent entities from those of non-equivalent ones, for each modality. In summary, our main contributions are:}

\begin{itemize}[itemsep=0.5ex, leftmargin=5mm]
\item 
\vic{We propose a framework to address the multi-modal entity alignment task by introducing multi-grained interaction mechanisms into the multi-modal knowledge representation process.}
%We propose a novel framework, called MMIEA, to address the multi-modal entity alignment task by introducing multi-grained interaction mechanisms in multi-modal knowledge representation process.
\item 
\vic{We design  mechanisms to explore intra-modal relationships and inter-modal interactions, ensuring that the aligned entities  are semantically close.} %with minimum gaps between modalities.} 
\item \vic{We conduct extensive experiments on two real-world datasets, showing the strong performance of MIMEA.}

%and superiority of MMIEA over the existing state-of-art models.}
\end{itemize}

\section{Related Work}
%\subsection{Entity Alignment}
\noindent \textbf{Entity Alignment.}
\vic{Entity alignment (EA), which aims to identify equivalent entities across different knowledge graphs, is a fundamental data integration task. Existing research focuses on learning embeddings of entities by utilizing the structural information of KGs. Approaches to EA  can be divided into two categories: \textit{KGE-based methods} and \textit{GNN-based methods}. KGE-based methods `move' entity embeddings from different KGs into a unified latent space and measure the alignment by calculating the distance between entity embeddings, such as MTransE~\citep{Muhao_2017}, JAPE~\citep{Zequn_2017}, IPTransE~\citep{Hao_2017}, BootEA~\citep{Zequn_2018}, RNM~\citep{Yao_2021} and NeoEA~\citep{Lingbing_2022}. Recently, GNN-based models have achieved remarkable performance in graph learning. Based on this, some works develop GNN-based frameworks for EA, such as KDCoE~\citep{Muhao_2018}, AliNet~\citep{Zequn_2020}, MuGNN~\citep{Yixin_2019}, AttrGNN~\citep{Zhiyuan_2020}. However, all the discussed methods ignore the multi-modal knowledge (especially the visual information) available in the knowledge graph. }
%Entity alignment (EA), which aims to identify equivalent entities across different knowledge graphs, is a fundamental data integration task. Existing researches focus on learning embeddings of entities by utilizing structureal information of KGs, which can be divided into two categories: \textit{KGE-based methods} and \textit{GNN-based methods}. The KGE-based methods transform entities embedding from different KGs into a unified latent space and measure the alignment by calculating the distance between entity embeddings, such as MTransE~\citep{Muhao_2017}, JAPE~\citep{Zequn_2017}, IPTransE~\citep{Hao_2017}, BootEA~\citep{Zequn_2018}, RNM~\citep{Yao_2021}, NeoEA~\citep{Lingbing_2022}. Recently, GNN-based models have achieved remarkable performance in graph learning, some works try to apply it to entity alignment for better representation learning, such as KDCoE~\citep{Muhao_2018}, AliNet~\citep{Zequn_2020}, MuGNN~\citep{Yixin_2019}, AttrGNN~\citep{Zhiyuan_2020}. However, the above methods ignore the multi-modal knowledge existing in the knowledge graph, especially the visual modality, and thus fail to fully utilize the multi-modal information to achieve better performance. 

\noindent \textbf{Multi-Modal Entity Alignment.} 
%\subsection{Multi-Modal Entity Alignment} 
\vic{Recently various multi-modal knowledge graphs have become available~\citep{Fangyu_2021, Ye_2019}. Thus many works have investigated how to effectively incorporate visual knowledge into the entity alignment task. PoE~\citep{Ye_2019} combines all multi-modal features into a single vector, and measures the trustworthiness of entity pairs by matching their underlying semantics. However, it cannot capture the potential interactions among different modalities. MMEA~\citep{Liyi_2020} integrates knowledge from different modalities into a joint representation and then calculates a similarity score between the holistic embeddings of aligned entities. EVA~\citep{Fangyu_2021} introduces an iterative learning strategy to expand the set of training seeds. HMEA~\citep{Hao_2021} encodes the multi-modal knowledge into the hyperbolic space, and uses aggregated embeddings to predict alignments. MSNEA~\citep{Liyi_2022} integrates visual features to guide the learning process of relation features and adaptively assigns attention weights to capture valuable attributes for alignment. MCLEA~\citep{Zhenxi_2022} explores intra-modal and inter-modal interactions via contrastive learning to reduce the gap between modalities. MEAformer~\citep{Zhuo_2022} proposes a transformer-based model which can dynamically predict relativized mutual weights among modalities for each entity, encoraging the emergence of adaptive modality preferences. ACK-MMEA~\citep{Qian_2023} designs a multi-modal attribute uniformization module to incorporate the consistent alignment knowledge. GEEA~\citep{Lingbing_2023} studies embedding-based entity alignment from a perspective of generative models. It converts an entity from one knowledge graph to the other one,  and generates new entities from random noise vectors.
However, the aforementioned methods have the following two shortcomings: On one hand, the majority of methods, such as MMEA, EVA, HMEA, and MSNEA, have not been able to fully achieve multi-granular interactions within and across modalities. Consequently, %the obtained entity embedding representations are suboptimal because 
they do not effectively integrate multimodal knowledge related to entities. On the other hand, even when some methods, like  MCLEA and  MEAformer, introduce mechanisms for intra-modality and inter-modality interactions, they are difficult to explicitly distinguish the importance of each modality, and also ignore the interaction between uni-modal and joint-modal embeddings, which results in a lack of diversity in modal interactions.} 

\noindent \vic{\textbf{Optimal Transport.} Optimal transport (OT) is a fundamental mathematical tool which aims  to derive an optimal plan to transfer one  distribution to another. OT  has been used in many applications, such as, computer vision~\citep{Nicolas_2011, Justin_2015}, domain adaption~\citep{Xiang_2022, Wanxing_2022}, and unsupervised learning~\citep{Mathilde_2020, Yuki_2020}. OTKGE~\citep{Zongsheng_2022} models the multi-modal fusion procedure as a transport plan moving different modal embeddings to a unified space by minimizing the Wasserstein distance between multi-modal distributions. MOTCat~\citep{Yingxue_2023} proposes a multi-modal optimal transport-based co-attention transformer framework with global structure consistency for selecting informative patches. However, existing studies lack a comprehensive investigation of the correlations between uni-modal and joint-modal contexts. To the best of our knowledge, we are the first to adopt the optimal transport mechanism for MMEA task.} 

\section{Preliminaries}
\noindent {\textbf{Multi-modal Knowledge Graph.}} 
\vic{Let $\mathcal{E}, \mathcal{R}, \mathcal{A}, \mathcal{I}, \mathcal{V}$ respectively be finite sets of entities, relation types, attribute types, images, and values. A \emph{multi-modal knowledge graph (MMKG)} $\mathcal{G}$ is defined as $\{\mathcal{P}, \mathcal{T}_r, \mathcal{T}_a\}$, where $\mathcal{P}=\{(e, i) \mid e \in \mathcal{E}, i \in \mathcal{I}\}$ is the set of \textit{entity-image} pairs, $\mathcal{T}_r=\{(h, r, t) \mid h, t \in \mathcal{E}, r \in \mathcal{R}\}$ is the set of \emph{relational triples}, and $\mathcal{T}_a=\{(e, a, v) \mid e \in \mathcal{E}, a \in \mathcal{A}, v \in \mathcal{V}\}$ is the set of \emph{attribute triples}.} 

% Let $\mathcal{E}, \mathcal{R}, \mathcal{A}, \mathcal{I}, \mathcal{V}$ respectively be finite sets of entities, relations, attributes, images, and values. A multi-modal knowledge graph (MMKG) is denoted as $\mathcal{G}=\{\mathcal{P}, \mathcal{T}_r, \mathcal{T}_a\}$, where $\mathcal{P}=\{(e, i) \mid e \in \mathcal{E}, i \in \mathcal{I}\}$ represents the set of \textit{entity-image} pairs, $\mathcal{T}_r=\{(h, r, t) \mid h, t \in \mathcal{E}, r \in \mathcal{R}\}$ refers to the set of relational triples, and $\mathcal{T}_a=\{(e, a, v) \mid e \in \mathcal{E}, a \in \mathcal{A}, v \in \mathcal{V}\}$ is the set of attribute triples.

\smallskip \noindent {\textbf{Multi-modal Entity Alignment.}} 
\vic{The aim of the \emph{multi-modal entity alignment (MMEA) task}  is to identify  pairs of entities in two multi-modal knowledge graphs which are equivalent. Concretely, given two MMKGs $\mathcal{G}=\{\mathcal{P}, \mathcal{T}_r, \mathcal{T}_a\}$ and $\mathcal{G}'=\{\mathcal{P}', \mathcal{T}_r', \mathcal{T}_a'\}$, we aim to find  entity pairs $\mathcal{H}=\{(e_i, e_j) \mid e_i \in \mathcal{E}, e_j \in \mathcal{E}', e_i \equiv e_j \}$, where $\equiv$ represents the equivalence of two entities. Usually, we will select a small set of pre-aligned entity pairs $\mathcal{S}$ (seeds) for training, to learn entity representations in the two input MMKGs.}

% The aim of multi-modal entity alignment (MMEA) is to identify whether a pair of entities in two multi-modal knowledge graphs is equivalent or not. Concretely, given two MMKGs $\mathcal{G}=\{\mathcal{P}, \mathcal{T}_r, \mathcal{T}_a\}$ and $\mathcal{G}'=\{\mathcal{P}', \mathcal{T}_r', \mathcal{T}_a'\}$, we need to find aligned entity pair $\mathcal{H}=\{(e_i, e_j) \mid e_i \in \mathcal{E}, e_j \in \mathcal{E}', e_i \equiv e_j \}$, where $\equiv$ represents the equivalence of two entities. Normally, we will select a small set of pre-aligned entity pairs $\mathcal{S}$ (seeds) for training guidance, to learn entity representations in two MMKGs.

\section{Framework}
\begin{figure*}[!htp]
    \centering
    \includegraphics[width=1\textwidth]{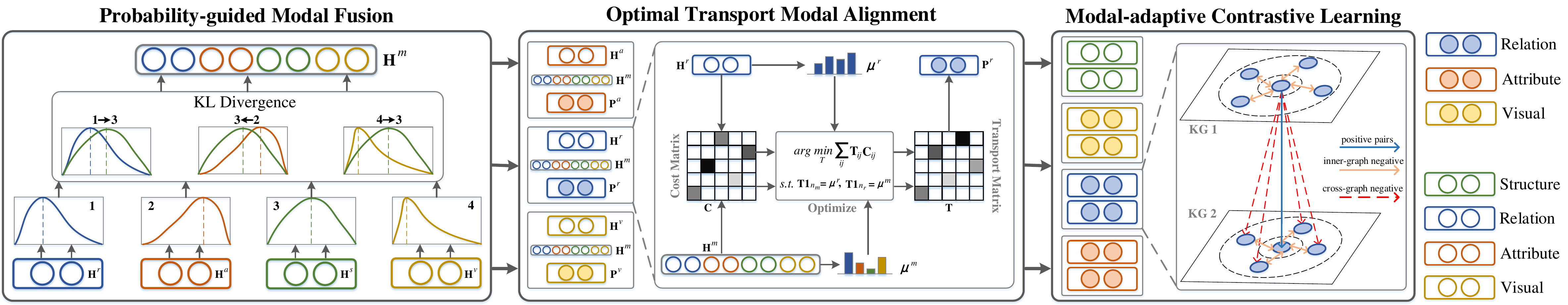}
    \caption{ MIMEA's architecture,  containing the modules: Probability-guided Modal Fusion, Optimal Transport Modal Alignment, and Modal-adaptive Contrastive Learning. }
    \label{figure_model}
\end{figure*}

\vic{We now introduce the \textbf{MIMEA} framework (cf.  Fig~\ref{figure_model} for its architecture), which comprises four major components (cf. Sections~\ref{sec:sec1}-\ref{sec:seclast}).}

%We propose a \textbf{M}ulti-Grained \textbf{M}ulti-Modal \textbf{I}nteraction framework for cross knowledge graph \textbf{E}ntity \textbf{A}lignment (\textbf{MMIEA}) - the  architecture is shown in Figure~\ref{figure_model}.  MMIEA comprises four major components, described in Sections~\ref{sec:sec1}-\ref{sec:seclast}.}

%: 1) \textit{Multi-modal Knowledge Embedding} (MKE) module to extract structure, relation, attribute, and visual modal with individual encoders; 2) \textit{Probability-guided Modal Fusion} (PMF) module to combine multi-modal knowledge to generate joint-modal representations; 3) \textit{Optimal Transport Modal Alignment} (OTMA) module to obtain the intermediate embedding between the uni-modal and joint-modal embeddings; 4) \textit{Modal-adaptive Contrastive Learning} (MCL) module to achieve the intra-modal interaction.
\subsection{Multi-modal Knowledge Embedding}\label{sec:sec1} We define entity embeddings for four modalities: structural, relation, attribute and visual. Structural embeddings are obtained based on the attribute and relational neighbors (described by attribute/relational triples)  of an entity. Relation embeddings are derived from  relation types, and they  are expressed in the form of  bag-of-words. Attribute embeddings are obtained analogously. Visual embeddings are derived from  entity-image pairs.
%\textcolor{green}{It should be noted that we consider four modalities of knowledge, namely structural modal, relation modal, attribute modal and visual modal. The structural modal is composed of relational triples and attribute triples. The relation modal is derived from the relational triples, but the relation modal is expressed in the form of bag-of-words, which determines whether the corresponding position of the corresponding embedding vector is 0 or 1 based on the relationship category in the relationship triples. The attribute modal is similar to the relation modal. The difference is that it is judged based on the attribute triples. The visual modal knowledge mainly comes from entity-image pairs.}

\smallskip \noindent\textbf{Structural Embeddings.} \vic{The graph attention network (GAT)~\citep{Petar_2018} is an attention-based architecture which has been shown to effectively encode graph-like data. We thus  leverage GAT to model the structural information of $\mathcal{G}$ and $\mathcal{G}'$. %More precisely, f
 For the hidden state $\textit{\textbf{h}}_i \in \mathbb{R}^d$ ($d$ represents the embedding dimension) of entity $e_i$, the aggregation of its one-hop neighbors $\mathcal{N}_i$ with self-loops is formulated as:
\begin{equation}
\textit{\textbf{h}}_i=\sigma \left(\sum_{j \in \mathcal{N}_i}{\alpha_{ij} \textbf{W}_s \textit{\textbf{h}}_j}\right)
\label{equation_1}
\end{equation}}
% The graph attention network (GAT) ~\citep{Petar_2018} calculated the interaction weight between a node and its neighbor nodes based on the attention mechanism, which can be better used to encode graph structure data, we leverage GAT to model the structural information of $\mathcal{G}$ and $\mathcal{G}'$. Specifically, for the hidden state $\textit{\textbf{h}}_i \in \mathbb{R}^d$ ($d$ represents the embedding dimension) of entity $e_i$ the aggregation of its one-hop neighbors $\mathcal{N}_i$ with self-loop is formulated as:
% \begin{equation}
% \textit{\textbf{h}}_i=\sigma \left(\sum_{j \in \mathcal{N}_i}{\alpha_{ij} \textbf{W}_s \textit{\textbf{h}}_j}\right)
% \label{equation_1}
% \end{equation} 
\vic{where $\sigma(\cdot)$ denotes the nonlinear ReLU  function; $\textbf{W}_s \in \mathbb{R}^{d \times d}$ denotes a parameterized weight matrix~\citep{Chengjiang_2019, Zhenxi_2022} --- we restrict $\textbf{W}_s$ to a diagonal matrix to reduce the number of  computations; $\textit{\textbf{h}}_j$ is the hidden state of entity $e_j$; the attention weight $\alpha_{ij}$ measures the importance  of entity $e_j$ for entity $e_i$, formulated as:}
% where $\sigma(\cdot)$ denotes the ReLU nonlinearity function, $\textbf{W}_s \in \mathbb{R}^{d \times d}$ denotes the parameterized weight matrix for shared linear transformation, similar as ~\citep{Chengjiang_2019, Zhenxi_2022}, we restrict $\textbf{W}_s$ to be a diagonal matrix to reduce computations. $\textit{\textbf{h}}_j$ is the hidden state of entity $e_j$, the attention weight $\alpha_{ij}$ measures the importance score of entity $e_j$ for  entity $e_i$, which can be formulated as:
\begin{equation}
\alpha_{ij}=\frac{\mathrm{exp} \left(\mathrm{LeakyReLU} \left(\textbf{a}^{\top}[\textbf{W}_s \textit{\textbf{h}}_i \parallel \textbf{W}_s \textit{\textbf{h}}_j]\right)\right)}{\sum_{k \in \mathcal{N}_i}{\mathrm{exp} \left(\mathrm{LeakyReLU} \left(\textbf{a}^{\top}[\textbf{W}_s \textit{\textbf{h}}_i \parallel \textbf{W}_s \textit{\textbf{h}}_k]\right)\right)}}
\label{equation_2}
\end{equation}
\vic{where $\textbf{a} \in \mathbb{R}^{2d}$ is a learnable parameter, $\cdot^{\top}$ and $\parallel$ respectively represent the  transposition and concatenation operations. To stabilize the learning process of self-attention, we introduce a multi-head strategy~\citep{Chengjiang_2019, Zhenxi_2022, Zhuo_2022} to generate \textit{K} independent representations based on the transformation of Equation ~\ref{equation_1}. Then, we concatenate these features to obtain the structural embedding $\textit{\textbf{h}}_i^s$ of entity $e_i$ as: }
%where $\textbf{a} \in \mathbb{R}^{2d}$ is a learnable parameter, $\cdot^{\top}$ represents transposition and $\parallel$ is the concatenation operation. To stabilize the learning process of self-attention, consistent with ~\citep{Chengjiang_2019, Zhenxi_2022, Zhuo_2022}, we introduce the multi-head strategy to generate \textit{K} independent representation based on the transformation of Equation ~\ref{equation_1}, then concatenate these features to obtain the structure embedding $\textit{\textbf{h}}_i^s$ of entity $e_i$ as following, and the structure embedding of all entities is represented as $\textbf{H}^s \in \mathbb{R}^{n \times d}$, $n$ represents the number of entities in the dataset:
\begin{equation}
\textit{\textbf{h}}_i^s=\mathop{\Vert}_{k=1}^{K}\sigma \left(\sum_{j \in \mathcal{N}_i}{\alpha_{ij}^{k} \textbf{W}_s^k \textit{\textbf{h}}_j}\right)
\label{equation_3}
\end{equation}
\vic{where $\alpha_{ij}^{k}$ denotes the normalized attention coefficients computed by the $k$-th attention mechanism, and $\textbf{W}_s^k$ is the corresponding input linear transformation's weight matrix. We use  a two-layer GAT to aggregate the neighborhood information across multiple hops, and use the output of the final GAT layer as the  structural embedding. The structural embedding of all entities is represented as $\textbf{H}^s \in \mathbb{R}^{n \times d}$, where $n$ represents the number of entities in the input dataset.}

% where $\alpha_{ij}^{k}$ denotes the normalized attention coefficients computed by the $k$-th attention mechanism, and $\textbf{W}_s^k$ is the corresponding input linear transformation's weight matrix. We perform a two-layer GAT to aggregate the neighborhood information across multiple hops, and use the output of final GAT layer as the neighbor-aware structure embedding.

%\subsubsection{\textbf{Relation and Attribute Embedding.}} 

\smallskip
\noindent \textbf{Relation and Attribute Embeddings.}
\vic{ Note that the knowledge from attribute types is coarser than that of relational types. Thus, directly mixing the representations of relations and attributes using a GAT can  easily  lead to the problem of information contamination~\citep{Fangyu_2021}. To alleviate this issue, we respectively regard the relations and attributes of entity $e_i$ as bag-of-words features $w_i^r$ and $w_i^a$. We further apply the multi-layer perceptrons $\mathrm{MLP}_r$ and $\mathrm{MLP}_a$ to respectively obtain the relation embedding $\textit{\textbf{h}}_i^r$ and attribute embedding $\textit{\textbf{h}}_i^a$,  calculated as:}
% Compared to relation triple knowledge, attribute triple knowledge is coarser-grained, mixing relations and attributes representation directly using GAT can easily lead to the problem of information contamination~\citep{Fangyu_2021}. To alleviate this issue, we regard the relations and attributes of entity $e_i$ as bag-of-words feature $w_i^r$ and $w_i^a$, respectively, further apply the separate multi-layer perceptron $\mathrm{MLP}_r$ and $\mathrm{MLP}_a$ to obtain the relation embedding $\textit{\textbf{h}}_i^r$ and attribute embedding $\textit{\textbf{h}}_i^a$, which can be calculated as follows, the relation and attribute embedding of all entities are represented as $\textbf{H}^r \in \mathbb{R}^{n \times d}$ and $\textbf{H}^a \in \mathbb{R}^{n \times d}$:
\begin{equation}
\textit{\textbf{h}}_i^r=\mathrm{MLP}_r(w_i^r), \quad \textit{\textbf{h}}_i^a=\mathrm{MLP}_a(w_i^a)
\label{equation_4}
\end{equation}
\vic{The relation and attribute embedding of all entities are respectively represented as $\textbf{H}^r \in \mathbb{R}^{n \times d}$ and $\textbf{H}^a \in \mathbb{R}^{n \times d}$}.

%\subsubsection{\textbf{Visual Embedding.}} 

\smallskip 
\noindent \textbf{Visual Embeddings.} \vic{VGG~\citep{Karen_2014} are usually pre-trained on large-scale image datasets and can extract useful features from images that are beneficial to different visual tasks. In practice, we feed the image $v_i$ of entity $e_i$ into the VGG-16 encoder $\mathrm{Enc}_v$. We use the final layer output before logits as the visual feature, and finally apply a multi-layer perceptron $\mathrm{MLP}_v$ to obtain the visual embedding $\textit{\textbf{h}}_i^v$:  }
\begin{equation}
\textit{\textbf{h}}_i^v=\mathrm{MLP}_v(\mathrm{Enc}_v(v_i))
\label{equation_5}
\end{equation}
\vic{The visual embedding of all entities is represented as $\textbf{H}^v \in \mathbb{R}^{n \times d}$.}

\subsection{Probability-guided Modal Fusion}
\vic{Different modalities concentrate on different types of knowledge. Thus, each modality  contributes differently to the characterization of specific aspects of an entity. Typically, it is  required to combine multiple modalities of knowledge to provide a more comprehensive understanding of an entity. For example, knowledge about the  entity \textit{Lionel Messi} includes  the relational triple  (\textit{Lionel Messi, employ, FC Barcelona}) and a visual image (an image of Messi wearing a certain team's jersey). So, when evaluating the football club \textit{Lionel Messi} plays for, the structural knowledge from the relational triple is more relevant than the  knowledge from the image. However, when it comes to Messi's jersey number at a club, the triple  (\textit{Lionel Messi, employ, FC Barcelona}) does not contain relevant information, but a visual image of Messi wearing a \textit{10} jersey  can provide  more appropriate clues. Holistically combining these two types of information will thus enable an accurate representation of the football club \textit{Lionel Messi} plays for and his jersey number at that club. Therefore, an important challenge is \textit{how to better integrate multi-modal knowledge to obtain effective  fused representations in multi-modal contexts.}}

\vic{A key source of knowledge in multi-modal knowledge graphs is the one provided by structural triples. The structural triples contain the relational triples and attribute triples, they can provide a more direct representation of the content of an entity and its relationship with other entities. For example, in the FB15K-DB15K dataset, there are a total of 714,720 structural triples, resulting in richer knowledge about the connections among entities.  In contrast, the FB15K-DB15K dataset contains only 1,624 relational types and 341 attribute types, which means that the initialization vectors for the relation and attribute modalities will be bag-of-words vectors of length 1,624 and 341. Consequently, the representation of relation  and attribute modalities of an entity lacks sufficient distinctiveness. Indeed, in subsequent ablation experiments we will show that the structural content has the most significant impact on the final performance of entity alignment. Therefore, using structural embeddings as a pivotal point, we introduce the \textit{Probability-guided Modal Fusion (PMF)} module, which employs a probabilistic distribution to achieve initial interactions between relation embeddings and structural embeddings, attribute embeddings and structural embeddings, also visual embeddings and structural embeddings. It generates interactive weights in the first stage and aggregates different modal embeddings to obtain a joint-modal combined representation based on these weight coefficients. Specifically, the PMF module comprises the following three steps:}
\begin{enumerate}[itemsep=0.5ex, leftmargin=5mm]
\item \textit{Constructing Probability Distributions.} 
\vic{Given the structural embedding $\textbf{H}^s$, relation embedding $\textbf{H}^r$, attribute embedding $\textbf{H}^a$, and visual embedding $\textbf{H}^v$ of all entities, we represent each embedding using a probability density form based on the Beta probability distribution function. The Beta distribution has two shape hyperparameters $\alpha$ and $\beta$. Its \emph{probability density function (PDF)} is defined as: $f_{(\alpha, \beta)}(x)=\frac{x^{\alpha-1}(1-x)^{\beta-1}}{\textbf{B}(\alpha, \beta)}$, where $x \in [0,1]$ and $\textbf{B}(\cdot)$ denotes the Beta function. To transform, for example,  the structural embedding $\textbf{H}^s$ into a Beta distribution, we proceed as follows: i) we first divide $\textbf{H}^s$ into two equal parts $\boldsymbol{\alpha}^s$ and $\boldsymbol{\beta}^s$ according to the embedding dimension:  $\textbf{H}^s=\{[\boldsymbol{\alpha}^s, \boldsymbol{\beta}^s] \, \vert \, \boldsymbol{\alpha}^s \in \mathbb{R}^{n \times m}, \boldsymbol{\beta}^s \in \mathbb{R}^{n \times m}, m=\frac{d}{2}\}$. Then, we use each part as a  shape parameter of the Beta distribution. ii) By combining  the \textit{i}-th element $\alpha^s_i$ in $\boldsymbol{\alpha}^s$ with the \textit{i}-th element $\beta^s_i$ in $\boldsymbol{\beta}^s$, we will form the \textit{i}-th Beta distribution. The combination of all elements will form \textit{m} Beta distributions, represented as $\mathcal{D}^s=[(\alpha^s_1, \beta^s_1), \ldots ,(\alpha^s_m, \beta^s_m)]$. We denote the PDF of the \textit{i}-th Beta distribution in $\mathcal{D}^s$ as $p_i^s$.} \vic{iii) We can analogously get the Beta distributions of  the relation, attribute and visual embedding: $\mathcal{D}^r$, $\mathcal{D}^a$, $\mathcal{D}^v$, and the corresponding \textit{i}-th Beta distributions: $p_i^r$, $p_i^a$, and $p_i^v$.}
\item \textit{Calculating Modal Weight Coefficients.} 
\vic{Given the relation and structural embedding's Beta distributions  $\mathcal{D}^r$ with parameters $[(\alpha^r_1, \beta^r_1),...,(\alpha^r_m, \beta^r_m)]$ and $\mathcal{D}^s$ with corresponding parameters $[(\alpha^s_1, \beta^s_1),...,(\alpha^s_m, \beta^s_m)]$, we define the distance between the relation and structural embedding as the sum of the KL divergence between the two Beta distributions along each dimension: }
% Regard for the relation embedding's Beta distributions $\mathcal{D}^r$ with parameters $[(\alpha^r_1, \beta^r_1),...,(\alpha^r_m, \beta^r_m)]$, and structure embedding's Beta distributions $\mathcal{D}^s$ with parameters $[(\alpha^s_1, \beta^s_1),...,(\alpha^s_m, \beta^s_m)]$, we define the distance between the relation and structure embedding as the sum of KL divergence between the two Beta embeddings along each dimension:
\begin{equation}
\delta_{(r,s)}=\sum_{i=1}^{m}{\mathrm{KL}(p_i^r, p_i^s)}
\label{equation_6}
\end{equation}
\vic{Then, we convert the KL distance $\delta_{(r,s)}$ to a weight coefficient based on $w_{(r,s)}=\lambda(2-\delta_{(r,s)})^2$, where $\lambda$ represents the incremental rate, set  empirically. Using the same method, we can obtain the weight coefficient $w_{(a,s)}$ between the attribute and structural embedding, and the weight coefficient $w_{(v,s)}$ between the visual and structural embedding. }

% Then we convert the KL distance $\delta_{(r,s)}$ to weight coefficient based on $w_{(r,s)}=\lambda(2-\delta_{(r,s)})^2$, where $\lambda$ represents the incremental rate set by empirical, using the same method, we can obtain the weight coefficient $w_{(a,s)}$ between attribute and structure embedding, and the weight coefficient $w_{(v,s)}$ between visual and structure embedding. 
\item \textit{Fusing Different Modal Embeddings.} 
\vic{We add the three weight coefficients $w_{(r,s)}$, $w_{(a,s)}$, $w_{(v,s)}$ with the initialized value 1.0 (initially we assume that all modalities  have the same weight coefficients) and normalize it to obtain the prior assumption: $W_{\text{PMF}}=\textit{softmax}([w_{(r,s)}, w_{(a,s)}, w_{(v,s)}] + 1.0$). Then, we multiply these weight coefficients with the embedding representation of the corresponding modality, and concate the multiplied results to obtain the final fused modality representation $\textbf{H}^m \in \mathbb{R}^{n \times 4d}$.}
% We add the three weight coefficients $w_{(r,s)}$, $w_{(a,s)}$, $w_{(v,s)}$ with initialized number 1.0 (initially we assume that all modals have the same weight coefficients) and normalize to obtain the prior assumption: $W_{PMF}=softmax([w_{(r,s)}, w_{(a,s)}, w_{(v,s)}] + 1.0)$. Then we multiply these weight coefficients with the embedding representation of the corresponding modality, and concate the multiplied results to obtain the final fused modality representation $\textbf{H}^m \in \mathbb{R}^{n \times 4d}$.
\end{enumerate}

\subsection{Optimal Transport Modal Alignment}
The PMF deals with various modalities of knowledge by combining multimodal information. However, due to the introduced noise during the fusion process, an optimal representation of an entity cannot be based only on joint-modal information. For instance, if we are interested in the football club in which \textit{Lionel Messi} plays, given the joint-modal embedding incorporating the relational triple (\textit{Lionel Messi, employ, FC Barcelona}), the attribute triple (\textit{Lionel Messi, number, 10}) and the visual image of Messi wearing the Barcelona jersey number 10, the knowledge provided by the attribute triple is regarded as noise, while the one provided by the relational triple  is useful information. Therefore, in some cases, while using joint-modal embeddings, we  need to retain the knowledge of individual embeddings for each modality to assess the extent to which a single modality represents an entity in a certain context. So, a natural question is \textit{how to achieve better interaction between uni-modal and joint-modal embeddings to cover the information gap of single modalities and reduce the  noise of joint-modal embeddings?}

\vic{{%Given the cost of each element pair, 
Optimal transport (OT) aims to transport the density distribution of a group of elements to that of another group with minimal total cost. To consider the correlations of uni-modal and joint-modal representations, we can regard uni-modal as one group elements and joint-modal as another group elements. The expectation is that the uni-modal and joint-modal elements have an appropriate correlation with minimal total transportation cost. To achieve this, } we first generate an intermediate transition matrix by aligning and optimizing uni-modal and joint-modal embeddings.  Subsequently, by combining uni-modal information with the generated intermediate transition matrix, we obtain  an enriched uni-modal embedding. The \textit{Optimal Transport Modal Alignment (OTMA)} module  consists of the following  steps:}

% Optimal transport mechanism provides an effective approach to obtain an intermediate modal embedding between single-modal and joint-modal embeddings. By aligning and optimizing single-modal and joint-modal embeddings, an intermediate transition matrix can be generated. Subsequently, combining the single-modal information with this intermediate transition matrix yields a compromised single-modal embedding. Based on this, we introduce the \textit{Optimal Transport Modal Alignment (OTMA)} module, which consists of the following three steps:
\begin{enumerate}[itemsep=0.5ex, leftmargin=5mm]
\item \textit{Building the  Transport Task.} \vic{We look, for instance, at how to obtain the intermediate modal embedding between the relation modal embedding $\textbf{H}^r$ and the joint modal embedding $\textbf{H}^m$. Optimal transport aims at computing a minimal cost transportation between a source distribution $\mu^r$ and a target distribution $\mu^m$:}
% Take obtaining the intermediate modal embedding between relation modal embedding $\textbf{H}^r$ and joint modal embedding $\textbf{H}^m$ as an example, optimal transport aims at computing a minimal cost transportation between a source distribution $\mu^r$, and a target distribution $\mu^m$, which can be written as:
\begin{equation}
\mu^r=\sum_{i=1}^{n_r}{q_i^r\varphi(x_i)}, \,\,\, \mu^m=\sum_{i=1}^{n_m}{q_i^m\varphi(y_i)}
\label{equation_7}
\end{equation}
\vic{where $\mu^r$ and $\mu^m$ are defined on the probability space $\textbf{H}^r$ and $\textbf{H}^m$, $\varphi(\cdot)$ denotes the Dirac function, $n_r$ and $n_m$ are the number of samples,  $x_i$ and $y_i$ are the \textit{i}-th sample of $\textbf{H}^r$ and $\textbf{H}^m$ (in practice, to reduce the computational complexity, the number of selected samples will be lower than the embedding dimension), $q_i^r$ and $q_i^m$ are the probability mass of the \textit{i}-th samples, satisfying the following conditions: $\sum_{i=1}^{n_r}{q_i^r}=\sum_{i=1}^{n_m}{q_i^m}=1$, to simplify the calculations, we set $q_i^r=\frac{1}{n_r}$ and $q_i^m=\frac{1}{n_m}$. We  define a cost matrix $\textbf{C}$ with $\textbf{C}_{ij}$ representing the distance (usually the cosine distance) between $x_i$ and $y_j$.}
% where $\mu^r$ and $\mu^m$ are defined on probability space $\textbf{H}^r$ and $\textbf{H}^m$, $\varphi(\cdot)$ denotes the Dirac function, $n_r$ and $n_m$ are the number samples, $x_i$ and $y_i$ are the \textit{i}-th sample of $\textbf{H}^r$ and $\textbf{H}^m$, in practice, in order to reduce computational complexity, the number of selected samples will be lower than the embedding dimension, $q_i^r$ and $q_i^m$ are the probability mass to the \textit{i}-th sample, satisfying the following conditions, i.e., $\sum_{i=1}^{n_r}{q_i^r}=\sum_{i=1}^{n_m}{q_i^m}=1$, to simplify the calculations, we set $q_i^r=\frac{1}{n_r}$ and $q_i^m=\frac{1}{n_m}$. We can define a cost matrix $\textbf{C}$ with $\textbf{C}_{ij}$ representing the distance between $x_i$ and $y_j$, normally, we will choose the  cosine distance.
\item \textit{Optimal Transport Plan.} 
\vic{Based on distributions $\mu^r$ and $\mu^m$, we can obtain all joint probability distributions $\Pi(\mu^r, \mu^m)$. Combining them with the cost matrix $\textbf{C}$, we can convert the optimal transport into the following form:}
% Based on distributions $\mu^r$ and $\mu^m$, we can obtain all joint probability distributions $\Pi(\mu^r, \mu^m)$, combined with the cost matrix $\textbf{C}$, we can convert the optimal transport into the following form:
\begin{equation}
\mathcal{W}(\mu^r, \mu^m)=\underset{\textbf{T} \in \Pi(\mu^r, \mu^m)}{\textit{min}} \sum_{i=1}^{n_r}\sum_{j=1}^{n_m}\textbf{T}_{ij}\textbf{C}_{ij}
\label{equation_8}
\end{equation}
\vic{where $\Pi(\mu^r, \mu^m)=\{\textbf{T} \in \mathbb{R}^{n_r\times n_m}|\textbf{T1}_{n_m}=\mu^r,\textbf{T}^{\top}\textbf{1}_{n_r}=\mu^m \}$, with  \textbf{1} an all-one vector, $\textbf{T}_{ij}$  the optimal amount of mass to move from $x_i$ to $y_j$ to obtain an overall minimum cost. We apply the Sinkhorn algorithm~\citep{Marco_2013} to optimize Equation~\eqref{equation_8} to get the optimal transportation matrix $\textbf{T}$.}
%where $\Pi(\mu^r, \mu^m)=\{\textbf{T} \in \mathbb{R}^{n_r\times n_m}|\textbf{T1}_{n_m}=\mu^r,\textbf{T}^{\top}\textbf{1}_{n_r}=\mu^m \}$, \textbf{1} denotes an all-one vector, $\textbf{T}_{ij}$ denotes the optimal amount of mass to move from $x_i$ to $y_j$ in order to obtain an overall minimum cost. We apply the Sinkhorn algorithm ~\citep{Marco_2013} to optimize Equation ~\ref{equation_8} to get the optimal transportation matrix $\textbf{T}$.
\item \textit{Translating Uni-Modal Embeddings.} \vic{We multiply the relation embedding $\textbf{H}^r$ with the transportation matrix $\textbf{T}$ to get the intermediate embedding $\textbf{P}^r={\textbf{H}^r}^{\top}\textbf{T}$ between the relation-modal embedding $\textbf{H}^r$ and the joint-modal embedding $\textbf{H}^m$. The resulting embedding  focuses on relational modal  knowledge, but can also be aligned with joint modal embeddings at a small cost.}
% We multiply the relation embedding $\textbf{H}^r$ with the transportation matrix $\textbf{T}$ to get the compromise embedding $\textbf{P}^r={\textbf{H}^r}^{\top}\textbf{T}$ between relation-modal embedding $\textbf{H}^r$ and joint-modal embedding $\textbf{H}^m$, which focuses on relational modal embedding knowledge, but can also be aligned with joint modal embeddings at a small cost. In the same way, we can obtain attribute and visual compromise embeddings, denoted as $\textbf{P}^a$ and $\textbf{P}^v$ respectively. We found that there is no need to align structure-modal embedding with joint-modal embedding. The reason is that the content structure embedding in joint embedding has the largest weight and dominates joint embedding.
\end{enumerate}
\vic{We can  analogously obtain attribute and visual intermediate embeddings, denoted as $\textbf{P}^a$ and $\textbf{P}^v$, respectively. We found that there is no need to align the structural-modal embedding with the joint-modal embedding since the  structural embedding in the joint embedding has the largest weight and therefore dominates the joint embedding.}

\subsection{Modal-adaptive Contrastive Learning}
\label{sec:seclast}
\vic{The OTMA module focuses on the interaction between uni-modal and joint-modal aspects of knowledge. However, both the OTMA and PMF modules overlook the interactions within a single modality. In many cases, for a given entity, there exist multiple associated pieces of information within a single modality. When predicting a specific attribute of an entity, typically only a subset of these related pieces of knowledge plays a decisive role. For instance, consider the entity \textit{Lionel Messi}, which includes the relational triples: (\textit{Lionel Messi, spouse, Antonela Roccuzzo}) and (\textit{Lionel Messi, child, Thiago Messi}) related to family relationships and (\textit{Lionel Messi, teammate, Neymar}) and (\textit{Lionel Messi, coach, Josep Guardiola}) related to player attributes. Clearly, when describing Messi's family relationships, \textit{Antonela Roccuzzo} and \textit{Thiago Messi} are  more important than \textit{Neymar} and \textit{Josep Guardiola}. However, when discussing Messi's football career, the situation is reversed. Therefore,  it is preferable to make the embeddings of \textit{Antonela Roccuzzo} and \textit{Thiago Messi} closer in the embedding space, while the embeddings of \textit{Antonela Roccuzzo} and \textit{Neymar} should be pushed farther apart. Based on these observations, an important challenge is \textit{how to enforce embeddings to respect modal properties, while distinguishing the embedding of an entity from those of other entities, for each modality.}}

\vic{Inspired by the contrastive learning mechanism~\citep{Mohammadreza_2021, Zhenxi_2022, Yanqiao_2021}, we devise  a \textit{Modal-adaptive Contrastive Learning (MCL)} module, which maps inner-graph aligned pairs to a proximate location, but  also pushes the inner-graph and cross-graph unaligned pairs father apart. Specifically, MCL includes the following three parts:}
%Inspired by the contrastive learning mechanism, we devise \textit{Modal-adaptive Contrastive Learning (MCL)} module, which can map inner-graph aligned pairs to a proximate location, but also make the inner-graph and cross-graph unaligned pair father apart. Specifically, MCL includes the following three parts:
\begin{itemize}[itemsep=0.5ex, leftmargin=5mm]
\item \textit{Creating Positive and Negative Samples.} \vic{Following a 1-to-1 alignment constraint~\citep{Zhenxi_2022,Zhuo_2022}, the entity pairs within the seed alignments $\mathcal{S}$ can be naturally regarded as positive samples, whereas any non-aligned pairs can be regarded as negative samples. \textcolor{black}{Let ($e_i^1, e_i^2$) in $\mathcal{S}$ (with $e_i^1 \in \mathcal{G}$ and $e_i^2 \in \mathcal{G}'$) be the \textit{i}-th aligned entity pair, the negative samples of $e_i^1$ are obtained from two sources: the inner-graph unaligned pairs from $\mathcal{G}$ and cross-graph unaligned pairs from $\mathcal{G}'$. More precisely, they are defined as $\mathcal{N}_i^1=\{e_j^1 ~|~ \forall e_j^1 \in \mathcal{G}, j \neq i\}$ and $\mathcal{N}_i^2=\{e_j^2 ~|~ \forall e_j^2 \in \mathcal{G}', j \neq i\}$.}
%For \textit{i}-th aligned entity pair ($e_i^1, e_i^2$) in $\mathcal{S}$  with $e_i^1 \in \mathcal{G}$ and $e_i^2 \in \mathcal{G}'$, for the entity $e_i^1$, the negative samples from two sources, inner-graph unaligned pairs from $\mathcal{G}$ and cross-graph unaligned pairs from $\mathcal{G}'$, which can be defined as $\mathcal{N}_i^1=\{e_j^1 \vert \forall e_j^1 \in \mathcal{G}, j \neq i\}$ and $\mathcal{N}_i^2=\{e_j^2 \vert \forall e_j^2 \in \mathcal{G}', j \neq i\}$.} 
It should be noted that we use the in-batch negative sampling strategy~\citep{Zhenxi_2022,Zhuo_2022} to limit the negative sample scope within the mini-batch.}

% \victor{Zhiwei, I can simply not parse the text in red. I tried but it is too convoluted. Could you please have a look.}
% Following the 1-to-1 alignment constraint~\citep{Zhenxi_2022,Zhuo_2022}, the entity pairs within the seed alignments $\mathcal{S}$ can be naturally regard as positive samples, whereas any non-aligned pairs can be regard as negative samples. For \textit{i}-th aligned entity pair ($e_i^1, e_i^2$) in $\mathcal{S}$, where $e_i^1 \in \mathcal{G}$ and $e_i^2 \in \mathcal{G}'$, for the entity $e_i^1$ in the ($e_i^1, e_i^2$), the negative samples from two sources, inner-graph unaligned pairs from $\mathcal{G}$ and cross-graph unaligned pairs from $\mathcal{G}'$, which can be defined as $\mathcal{N}_i^1=\{e_j^1 \vert \forall e_j^1 \in \mathcal{G}, j \neq i\}$ and $\mathcal{N}_i^2=\{e_j^2 \vert \forall e_j^2 \in \mathcal{G}', j \neq i\}$. It should be noted that we introduce the in-batch negative sampling strategy~\citep{Zhenxi_2022,Zhuo_2022} to limit the negative sample scope within the mini-batch for efficiency.
\item \textit{Contrastive Learning Loss.} \vic{For the constructed positive and negative examples, we  perform contrastive learning under each modal condition. For instance, for the relational modality, we  construct the contrastive learning loss $\mathcal{L}^r(e_i^1, e_i^2)$ of the positive pair ($e_i^1, e_i^2$) as:}
% For the constructed positive and negative examples, we can perform contrastive learning under each modal condition. Taking the relational modality as an example, we can construct the contrastive learning loss of positive pair ($e_i^1, e_i^2$) as:
\begin{equation}
 -\mathrm{log}\frac{\theta(e_i^1, e_i^2)}{\theta(e_i^1, e_i^2) + \gamma \, \sum_{e_j^1 \in \mathcal{N}_i^1}{\theta(e_i^1, e_j^1)} + \sum_{e_j^2 \in \mathcal{N}_i^2}{\theta(e_i^1, e_j^2)}}
\label{equation_9}
\end{equation}
\vic{where $\theta(x,y)=exp(f_r(x)^{\top}f_r(y)/\tau)$, $f_r(\cdot)$ is the relation encoder, $\tau$ is a temperature parameter, and $\gamma$ is a hyper-parameter to control inner-graph alignment. The second and third terms in the denominator sum up inner-graph and cross-graph negative samples, respectively. We apply L2-normalisation to the input feature embeddings before computing the inner product~\citep{Mohammadreza_2021, Zhenxi_2022, Yumin_2019}. Similarly, we can obtain the  loss  for the other direction as $\mathcal{L}^r(e_i^2, e_i^1)$. The final contrastive loss of the relational modality is the average of the losses in the two directions, expressed as: $\mathcal{L}^r=\frac{1}{2}[\mathcal{L}^r(e_i^1, e_i^2)+\mathcal{L}^r(e_i^2, e_i^1)]$.}
%where $\theta(x,y)=exp(f_r(x)^{\top}f_r(y)/\tau)$, $f_r(\cdot)$ is the relation encoder, $\tau$ is a temperature parameter, $\gamma$ is a hyper-parameter to control inner-graph alignment. The second and third term in the denominator sum up inner-graph and cross-graph negative samples, respectively. 
%We apply L2-normalisation to the input feature embeddings before computing the inner product~\citep{Mohammadreza_2021, Zhenxi_2022, Yumin_2019}. Similarly, we can obtain the other direction loss as $\mathcal{L}^r(e_i^2, e_i^1)$, and the final contrastive loss of relation modal is the average of the losses in the two directions, expressed as: $\mathcal{L}^r=\frac{1}{2}[\mathcal{L}^r(e_i^1, e_i^2)+\mathcal{L}^r(e_i^2, e_i^1)]$.
\item \textit{Optimization Objective.} \vic{Using the same idea, we can obtain the contrastive loss of structural, attribute, visual and joint modalities, respectively expressed as $\mathcal{L}^s$, $\mathcal{L}^a$, $\mathcal{L}^v$, and $\mathcal{L}^m$. The overall loss is defined as:}
%Using the same idea, we can obtain the contrastive loss of structure, attribute, visual and joint modality, respectively expressed as $\mathcal{L}^s$, $\mathcal{L}^a$, $\mathcal{L}^v$, and $\mathcal{L}^m$, and the overall loss of can be defined as follows:
\begin{equation}
\mathcal{L} = \sum_{\ell \in \mathcal{M}}{\phi^\ell \mathcal{L}^\ell}, \,\, \mathcal{M}=\{s, r, a, v, m\}
\label{equation_10}
\end{equation}
\vic{where $\phi^\ell$ is the hyper-parameter that balances the importance of different modal losses. Similar to~\citep{Alex_2018}, we introduce a multi-task learning paradigm and then use homoscedastic uncertainty to weight each loss automatically during model training. Details of this strategy can be found in~\citep{Alex_2018}. It should be noted that only the MCL module  has loss values, and the PMF and OTMA modules  do not have any loss content.}
%where $\phi^l$ is the hyper-parameter that balance the importance of different modal's losses. Similar as the ~\citep{Alex_2018}, we introduce a multi-task learning paradigm and then use homoscedastic uncertainty to weight each loss automatically during model training, details of this strategy can be found in the original paper. It should be noted that only module MCL has loss values, and modules PMF and OTMA do not have any loss content.
\end{itemize}

\section{Experiments}
\begin{table*}[!htp]
\setlength{\abovecaptionskip}{0.05cm}
\renewcommand\arraystretch{1.3}
\setlength{\tabcolsep}{0.25em}
\centering
\small
\caption{Evaluation of different models in the non-iterative setting. 
Results marked with $\dagger$, $\ddagger$ and $\ast$ respectively come from~\citep{Zhenxi_2022}~\citep{Zhuo_2022}, and  the corresponding paper. Best scores are  in \textbf{bold}, the second best scores are \uline{underlined}, and `--' indicates the results are not reported in previous work.}
\begin{tabular*}{0.99\linewidth}{@{}ccccccccccccccccccc@{}}
\bottomrule
\multicolumn{1}{c}{\multirow{3}{*}{\textbf{Methods}}} & \multicolumn{9}{c}{\textbf{FB15K-DB15K}} & \multicolumn{9}{c}{\textbf{FB15K-YAGO15K}}\\
\cline{2-7}\cline{8-13}\cline{14-19}
& \multicolumn{3}{c}{\textbf{20\%}} & \multicolumn{3}{c}{\textbf{50\%}} & \multicolumn{3}{c}{\textbf{80\%}} & \multicolumn{3}{c}{\textbf{20\%}} & \multicolumn{3}{c}{\textbf{50\%}} & \multicolumn{3}{c}{\textbf{80\%}} \\
\cline{2-3}\cline{4-5}\cline{6-7}\cline{8-9}\cline{10-11}\cline{12-13}\cline{14-15}\cline{16-17}\cline{18-19}

& \textbf{MRR} & \textbf{H@1} & \textbf{H@10} & \textbf{MRR}  & \textbf{H@1} & \textbf{H@10} & \textbf{MRR} & \textbf{H@1} & \textbf{H@10}  & \textbf{MRR} & \textbf{H@1} & \textbf{H@10} & \textbf{MRR}  & \textbf{H@1} & \textbf{H@10} & \textbf{MRR}  & \textbf{H@1} & \textbf{H@10}\\

\hline
PoE~\citep{Ye_2019}$^\dagger$ &0.170 &0.126 &0.251 &0.533 &0.464 &0.658 &0.721    &0.666 &0.820 &0.154 &0.113 &0.229 &0.414 &0.347 &0.536 &0.635 &0.573 &0.746 \\
HMEA~\citep{Hao_2021}$^\dagger$ &-- &0.127 &0.369 &-- &0.262 &0.581 &-- &0.417 &0.786 &-- &0.105 &0.313 &-- &0.265 &0.581 &-- &0.433 &0.801 \\
MMEA~\citep{Liyi_2020}$^\dagger$ &0.357 &0.265 &0.541 &0.512 &0.417 &0.703 &0.685    &0.590 &0.869 &0.317 &0.234 &0.480 &0.486 &0.403 &0.645 &0.682 &0.598 &0.839 \\
EVA~\citep{Fangyu_2021}$^\ddagger$ &0.283 &0.199 &0.448 &0.422 &0.334 &0.589 &0.563    &0.484 &0.696 &0.224 &0.153 &0.361 &0.388 &0.311 &0.534 &0.565 &0.491 &0.692 \\
MSNEA~\citep{Liyi_2022}$^\ddagger$ &0.175 &0.114 &0.296 &0.388 &0.288 &0.590 &0.613    &0.518 &0.779 &0.153 &0.103 &0.249 &0.413 &0.320 &0.589 &0.620 &0.531 &0.778 \\
MCLEA~\citep{Zhenxi_2022}$^\ddagger$ &0.393 &0.295 &0.582 &0.637 &0.555 &0.784 &0.790    &0.735 &0.890 &0.332 &0.254 &0.484 &0.574 &0.501 &0.705 &0.722 &0.667 &0.824 \\
MEAformer~\citep{Zhuo_2022}$^\ddagger$ &\uline{0.518} &\uline{0.417} &\uline{0.715} &0.698 &0.619 &0.843 &0.820 &0.765 &\uline{0.916} &\uline{0.417} &\uline{0.327} &\uline{0.595} &0.639 &0.560 &0.778 &0.766 &0.703 &0.873 \\
ACK-MMEA~\citep{Qian_2023}$^\ast$  &0.387 &0.304 &0.549 &0.624 &0.560 &0.736 &0.752    &0.682 &0.874 &0.360 &0.289 &0.496 &0.593 &0.535 &0.699 &0.744 &0.676 &0.864 \\
GEEA~\citep{Lingbing_2023}$^\ast$ &0.450 &0.343 &0.661 &\uline{0.723} &\uline{0.651} &\uline{0.852} &\uline{0.836} &\uline{0.787} &\textbf{0.918} &0.393 &0.298 &0.585 &\uline{0.668} &\uline{0.589} &\uline{0.808} &\uline{0.790} &\uline{0.733} &\textbf{0.890} \\
\hline
MIMEA  &\textbf{0.594}  &\textbf{0.506}  &\textbf{0.756}  &\textbf{0.748}  &\textbf{0.683}  &\textbf{0.861}  &\textbf{0.841}  &\textbf{0.799}  &0.914 &\textbf{0.506} &\textbf{0.417} &\textbf{0.671} &\textbf{0.692}   &\textbf{0.622} &\textbf{0.818}   &\textbf{0.795} &\textbf{0.741}   &\uline{0.884} \\
\bottomrule
\end{tabular*}
\label{table_non_iterative}
\end{table*}

\vic{To evaluate the effectiveness of MIMEA,  we aim to explore the following research questions:}
%To evaluate the effectiveness of MMIEA, in this section, we aim to answer the following research questions:
\begin{itemize}[itemsep=0.5ex, leftmargin=5mm]
\item 
\textbf{RQ1 (Effectiveness):} How does MIMEA perform compared to the SoTA?
\item 
\textbf{RQ2 (Ablation studies):} How do different components of MIMEA contribute to its performance? 
\item
\textbf{RQ3 (Complexity analysis):} 
What is the amount of computation and parameters used by MIMEA? 
\item 
\textbf{RQ4 (Parameter analysis):} How do hyper-parameters influence the performance of MIMEA? A detailed analysis can be found in Appendix \hyperlink{parameter_analysis}{D}.
\end{itemize}

\subsection{Experimental Setup}
%\subsubsection
\noindent \textbf{Datasets.} 
\vic{We evaluate the  MIMEA model on two well-known datasets: FB15K-DB15K and FB15K-YAGO15K, which include 12,846 and 11,199 alignment pairs, respectively. As in previous works~\citep{Zhuo_2022, Zhenxi_2022, Qian_2023}, to evaluate  MIMEA's performance under different conditions, we split the two datasets into training and testing sets with 20\%, 50\%, and 80\% of  pre-aligned pairs  given as  alignment seeds. The statistics of these datasets can be found in Appendix \hyperlink{datasets}{A}.}

%We evaluate the  MMIEA model on two well-known datasets, i.e., FB15K-DB15K and FB15K-YAGO15K, which includes 12,846 and 11,199 alignment seeds. As previous works~\citep{Zhuo_2022, Zhenxi_2022, Qian_2023}, to evaluate the MMIEA's performance under different conditions, we split the two datasets into training and testing sets according to 20\%, 50\%, and 80\% of pre-aligned pairs as the seed alignments, respectively. The statistics of corresponding datasets are shown in Appendix \hyperlink{datasets}{A}.

%\subsubsection{Iterative Training.} 
\smallskip
\noindent \textbf{Iterative Training.} 
\vic{As in previous works~\citep{Fangyu_2021, Liyi_2022, Zhenxi_2022, Zhuo_2022}, we adopt a probation strategy for \textit{iterative training}. Specifically, we constructed a buffer to temporarily store entity pairs that are close in the embedding space across different knowledge graphs. In every  round $R$, we select entity pairs that meet the nearest distance  
criteria and add them to the buffer. If after  $M$  iterations, these entity pairs are still  in the buffer, we will add them to the training set. This approach effectively serves as a data augmentation strategy during training, where the entity pairs in the buffer can be considered as pseudo-labels. %We need to continuously iterate to obtain entity pairs from the pseudo-labels that might be correct labels. 
In contrast, the training method that does not involve the aforementioned iterative process is referred to as \textit{non-iterative training}.}

%As in previous works~\citep{Fangyu_2021, Liyi_2022, Zhenxi_2022, Zhuo_2022}, we adopt a probation strategy for \textit{iterative training}. Specifically, we have constructed a buffer to temporarily store entity pairs that are close in embedding space across knowledge graphs. Every $R$ rounds, we select entity pairs that meet nearest criteria and add them to the buffer. If after the subsequent $M$ rounds of iterations, these entity pairs still exist in the buffer, we will add them to the training set. This approach effectively serves as a data augmentation strategy during training, where the entity pairs in the buffer can be considered as pseudo-labels. We need to continuously iterate to obtain entity pairs from the pseudo-labels that might be correct labels. In contrast, the training method that does not involve the aforementioned iterative process is referred to \textit{non-iterative training}.

%\subsubsection{Baselines.} 
\smallskip 
\noindent \textbf{Baselines.}
\vic{In the experiments, we used two training strategies: \textit{non-iterative} and \textit{iterative training}. For each training strategy, we used different baselines. For non-iterative training: PoE~\citep{Ye_2019}, HMEA~\citep{Hao_2021}, MMEA~\citep{Liyi_2020}, EVA~\citep{Fangyu_2021}, MSNEA~\citep{Liyi_2022}, MCLEA~\citep{Zhenxi_2022}, MEAformer~\citep{Zhuo_2022}, ACK-MMEA~\citep{Qian_2023}, and GEEA~\citep{Lingbing_2023}.
For iterative training: EVA~\citep{Fangyu_2021}, MSNEA~\citep{Liyi_2022}, MCLEA~\citep{Zhenxi_2022}, and MEAformer~\citep{Zhuo_2022}. Implementation details and evaluation metrics   can respectively be found in Appendix \hyperlink{hyper-parameters}{B} and \hyperlink{evaluation_metrics}{C}.}

\subsection{Main Results}
\vic{To address \textbf{RQ1}, we conduct experiments  on the  non-iterative and iterative training settings, and on the number of selected pre-aligned seeds. The results are shown in Tables~\ref{table_non_iterative} and~\ref{table_iterative}. }

%To answer \textbf{RQ1}, we evaluate MMIEA and conduct experiments on two benchmark datasets under non-iterative and iterative training settings, and select pre-aligned seeds to be 20\%, 50\%, and 80\%, respectively. The results can be show in Table~\ref{table_non_iterative} and Table~\ref{table_iterative}.
%\subsubsection{Performance Comparison.} 

\smallskip
\noindent \textbf{Performance Comparison.}
\vic{We can observe in the results that under both the non-iterative and iterative settings, MIMEA  generally outperforms existing SoTA baselines by a large margin across all metrics. More precisely, we have the following observations. On the one hand, MIMEA achieves the best performance on the multi-modal entity alignment task. %\zhiwei{as shown in Table~\ref{table_non_iterative} and Table~\ref{table_iterative}}\nb{cf. Table bla}. 
For example, in the  non-iterative setting, on the FB15K-YAGO15K dataset, MIMEA achieves improvements of 8.9\%, 2.4\% and  0.5\%  on MRR compared to the best SoTA baselines  when  the given pre-aligned seeds are  20\%, 50\%, and 80\%, respectively. %compared to the best performing baseline MEAformer or GEEA. 
Similar improvements are obtained  on the FB15K-DB15K dataset. On the other hand, the iterative training strategy can significantly improve model performance of  existing  baselines and MIMEA. For example, on the FB15K-DB15K dataset when the given pre-aligned seeds are 20\%, 50\%, and 80\%, depending on whether  MIMEA uses the iterative training mechanism, there will be fluctuations of  10\%, 2.2\%, and 1.4\% on MRR, respectively. This is primarily attributed to the  generation of pseudo-entity alignments pairs during the iterative training process, which iteratively  filters out potentially wrong entity pairs.}

\begin{table*}[!htp]
\setlength{\abovecaptionskip}{0.05cm}
\renewcommand\arraystretch{1.3}
\setlength{\tabcolsep}{0.27em}
\centering
\small
\caption{Evaluation of different models under iterative setting. $\ddagger$ results come from ~\citep{Zhuo_2022}. $\ast$ results from the corresponding papers. Best scores are highlighted in \textbf{bold}, the second best scores are \uline{underlined}.}
\begin{tabular*}{0.995\linewidth}{@{}ccccccccccccccccccc@{}}
\bottomrule
\multicolumn{1}{c}{\multirow{3}{*}{\textbf{Methods}}} & \multicolumn{9}{c}{\textbf{FB15K-DB15K}} & \multicolumn{9}{c}{\textbf{FB15K-YAGO15K}}\\
\cline{2-7}\cline{8-13}\cline{14-19}
& \multicolumn{3}{c}{\textbf{20\%}} & \multicolumn{3}{c}{\textbf{50\%}} & \multicolumn{3}{c}{\textbf{80\%}} & \multicolumn{3}{c}{\textbf{20\%}} & \multicolumn{3}{c}{\textbf{50\%}} & \multicolumn{3}{c}{\textbf{80\%}} \\
\cline{2-3}\cline{4-5}\cline{6-7}\cline{8-9}\cline{10-11}\cline{12-13}\cline{14-15}\cline{16-17}\cline{18-19}

& \textbf{MRR} & \textbf{H@1} & \textbf{H@10} & \textbf{MRR}  & \textbf{H@1} & \textbf{H@10} & \textbf{MRR} & \textbf{H@1} & \textbf{H@10}  & \textbf{MRR} & \textbf{H@1} & \textbf{H@10} & \textbf{MRR}  & \textbf{H@1} & \textbf{H@10} & \textbf{MRR}  & \textbf{H@1} & \textbf{H@10}\\

\hline
EVA~\citep{Fangyu_2021}$^\ddagger$ &0.318 &0.231 &0.488 &0.449 &0.364 &0.606 &0.573    &0.491 &0.711 &0.260 &0.188 &0.403 &0.404 &0.325 &0.560 &0.572 &0.493 &0.695 \\
MSNEA~\citep{Liyi_2022}$^\ddagger$ &0.232 &0.149 &0.392 &0.459 &0.358 &0.656 &0.651    &0.565 &0.810 &0.210 &0.138 &0.346 &0.472 &0.376 &0.646 &0.668 &0.593 &0.806 \\
MCLEA~\citep{Zhenxi_2022}$^\ast$ &0.534 &0.445 &0.705 &0.652 &0.573 &0.800 &0.784    &0.730 &0.883 &0.474 &0.388 &0.641 &0.616 &0.543 &0.759 &0.715 &0.653 &0.835 \\
MEAformer~\citep{Zhuo_2022}$^\ddagger$ &\uline{0.661} &\uline{0.578} &\uline{0.812} &\uline{0.755} &\uline{0.690} &\uline{0.871} &\uline{0.834} &\uline{0.784} &\textbf{0.921} &\uline{0.529} &\uline{0.444} &\uline{0.692} &\uline{0.682} &\uline{0.612} &\uline{0.808} &\uline{0.783} &\uline{0.724} &\uline{0.880} \\
\hline
MIMEA &\textbf{0.694} &\textbf{0.622} &\textbf{0.824} &\textbf{0.770} &\textbf{0.716} &\textbf{0.872} &\textbf{0.855} &\textbf{0.821} &\uline{0.919} &\textbf{0.587} &\textbf{0.513} &\textbf{0.729} &\textbf{0.712} &\textbf{0.651} &\textbf{0.827} &\textbf{0.803} &\textbf{0.757} &\textbf{0.885} \\
\bottomrule
\end{tabular*}
\label{table_iterative}
\end{table*}

\begin{table*}[!htp]
\setlength{\abovecaptionskip}{0.05cm}
\renewcommand\arraystretch{1.3}
\setlength{\tabcolsep}{0.30em}
\centering
\small
\caption{Ablation studies under different modals and different modules. Best scores are highlighted in \textbf{bold}.}
\begin{tabular*}{0.99\linewidth}{@{}ccccccccccccccccccc@{}}
\bottomrule
\multicolumn{1}{c}{\multirow{3}{*}{\textbf{Settings}}} & \multicolumn{9}{c}{\textbf{FB15K-DB15K}} & \multicolumn{9}{c}{\textbf{FB15K-YAGO15K}}\\
\cline{2-7}\cline{8-13}\cline{14-19}
& \multicolumn{3}{c}{\textbf{20\%}} & \multicolumn{3}{c}{\textbf{50\%}} & \multicolumn{3}{c}{\textbf{80\%}} & \multicolumn{3}{c}{\textbf{20\%}} & \multicolumn{3}{c}{\textbf{50\%}} & \multicolumn{3}{c}{\textbf{80\%}} \\
\cline{2-3}\cline{4-5}\cline{6-7}\cline{8-9}\cline{10-11}\cline{12-13}\cline{14-15}\cline{16-17}\cline{18-19}

& \textbf{MRR} & \textbf{H@1} & \textbf{H@10} & \textbf{MRR}  & \textbf{H@1} & \textbf{H@10} & \textbf{MRR} & \textbf{H@1} & \textbf{H@10}  & \textbf{MRR} & \textbf{H@1} & \textbf{H@10} & \textbf{MRR}  & \textbf{H@1} & \textbf{H@10} & \textbf{MRR}  & \textbf{H@1} & \textbf{H@10}\\

\hline
w/o structure &0.094 &0.044 &0.183 &0.134 &0.066 &0.264 &0.220 &0.117 &0.447 &0.088 &0.051 &0.151 &0.105 &0.056 &0.192 &0.175 &0.094 &0.335 \\
w/o attribute &0.664 &0.589 &0.806 &0.750 &0.694 &0.859 &0.839 &0.801 &0.910 &0.553 &0.479 &0.700 &0.684 &0.620 &0.806 &0.774 &0.718 &0.871 \\
w/o relation &0.642 &0.565 &0.785 &0.742 &0.685 &0.850 &0.831 &0.791 &0.902 &0.507 &0.429 &0.660 &0.661 &0.587 &0.801 &0.776 &0.717 &0.875 \\
w/o visual &0.691 &0.616 &\textbf{0.825} &\textbf{0.772} &\textbf{0.716} &\textbf{0.877} &0.853 &0.815 &0.921 &0.568 &0.492 &0.717 &0.699 &0.633 &0.822 &0.791 &0.736 &0.886 \\
\cdashline{1-19}
w/o PMF &0.595 &0.507 &0.757 &0.747 &0.682 &0.862 &0.841 &0.797 &0.914 &0.494 &0.406 &0.659 &0.688 &0.617 &0.817 &0.794 &0.741 &0.885 \\
w/o OTMA &0.576 &0.486 &0.748 &0.724 &0.648 &0.860 &0.845 &0.797 &\textbf{0.923} &0.518 &0.437 &0.673 &0.660 &0.578 &0.810 &0.796 &0.737 &\textbf{0.897} \\
w/o MCL &0.630 &0.535 &0.797 &0.744 &0.671 &0.873 &0.844 &0.802 &0.918 &0.535 &0.449 &0.690 &0.680 &0.612 &0.795 &0.780 &0.723 &0.878 \\
\hline
MIMEA &\textbf{0.694} &\textbf{0.622} &0.824 &0.770 &\textbf{0.716} &0.872 &\textbf{0.855} &\textbf{0.821} &0.919 &\textbf{0.587} &\textbf{0.513} &\textbf{0.729} &\textbf{0.712} &\textbf{0.651} &\textbf{0.827} &\textbf{0.803} &\textbf{0.757} &0.885 \\
\bottomrule
\end{tabular*}
\label{table_different_module}
\end{table*}

\smallskip \noindent \textbf{Impact of Number of Pre-aligned Seeds.} \vic{We evaluate the sensitivity of MIMEA to the given number of pre-aligned seeds: 20\%, 50\%, and 80\%~\citep{Fangyu_2021, Liyi_2022, Zhenxi_2022, Zhuo_2022}. From the results, %(\zhiwei{cf. Table~\ref{table_non_iterative} and Table~\ref{table_iterative}}\nb{cf. Table bla}), 
we can observe that MIMEA achieves the best performance on both the FB15K-DB15K and the FB15K-YAGO15K datasets in all metrics and proportions, confirming its robustness to the number of given pre-aligned seeds. For instance, in the iterative setting, on FB15K-YAGO15K, compared with the best-performing baseline MEAformer, for  20\%, 50\%, and 80\%, the MRR metric is respectively improved by 5.8\%, 3.0\% and 2.0\%. The higher improvement for 20\% shows that MIMEA is well-suited  for low-resource scenarios. 
\textcolor{black}{This is mainly because, on the one hand, each modality can be explicitly given a differentiation weight according to the characteristics of such modality. Further, we take into account the interactions between uni-modal and joint-modal representations. On the other hand, intra-modal is able to differentiate  uni-modal representations. The intra-modal and inter-modal multi-granularity interaction  can indeed maximize the utility of having multi-modal knowledge.}}

\subsection{Ablation Studies}
\vic{We address \textbf{RQ2} from four perspectives, including different variants, different modalities, different  distribution methods, and different pivotal modality. The results are shown in Tables~\ref{table_different_module}, ~\ref{table_different_distribution}, and~\ref{table_different_core}.}

\begin{table*}[!htp]
\setlength{\abovecaptionskip}{0.05cm}
\renewcommand\arraystretch{1.3}
\setlength{\tabcolsep}{0.33em}
\centering
\small
\caption{Evaluation of different models under different distribution methods. Best scores are highlighted in \textbf{bold}.}
\begin{tabular*}{0.99\linewidth}{@{}ccccccccccccccccccc@{}}
\bottomrule
\multicolumn{1}{c}{\multirow{3}{*}{\textbf{Settings}}} & \multicolumn{9}{c}{\textbf{FB15K-DB15K}} & \multicolumn{9}{c}{\textbf{FB15K-YAGO15K}}\\
\cline{2-7}\cline{8-13}\cline{14-19}
& \multicolumn{3}{c}{\textbf{20\%}} & \multicolumn{3}{c}{\textbf{50\%}} & \multicolumn{3}{c}{\textbf{80\%}} & \multicolumn{3}{c}{\textbf{20\%}} & \multicolumn{3}{c}{\textbf{50\%}} & \multicolumn{3}{c}{\textbf{80\%}} \\
\cline{2-3}\cline{4-5}\cline{6-7}\cline{8-9}\cline{10-11}\cline{12-13}\cline{14-15}\cline{16-17}\cline{18-19}

& \textbf{MRR} & \textbf{H@1} & \textbf{H@10} & \textbf{MRR}  & \textbf{H@1} & \textbf{H@10} & \textbf{MRR} & \textbf{H@1} & \textbf{H@10}  & \textbf{MRR} & \textbf{H@1} & \textbf{H@10} & \textbf{MRR}  & \textbf{H@1} & \textbf{H@10} & \textbf{MRR}  & \textbf{H@1} & \textbf{H@10}\\

\hline
Beta &0.694 &0.622 &0.824 &\textbf{0.770} &\textbf{0.716} &0.872 &\textbf{0.855} &\textbf{0.821} &0.919 &0.587 &0.513 &0.729 &0.712 &\textbf{0.651} &0.827 &0.803 &\textbf{0.757} &0.885 \\
Cauchy  &\textbf{0.697}  &\textbf{0.625}  &\textbf{0.825}  &\textbf{0.770}  &0.713  &\textbf{0.877}  &0.852  &0.815  &\textbf{0.921} &\textbf{0.592} &\textbf{0.516} &\textbf{0.734} &\textbf{0.715}   &\textbf{0.651} &\textbf{0.833}   &\textbf{0.806} &\textbf{0.757}   &0.891 \\
Gamma &0.691 &0.621 &0.822 &0.769 &0.713 &0.874 &0.851 &0.818 &0.915 &0.575 &0.496 &0.724 &0.708 &0.646 &0.828 &0.802 &0.751 &\textbf{0.893} \\
Gumbel &0.690 &0.619 &0.821 &0.769 &0.714 &0.872 &0.851 &0.816 &0.917 &0.573 &0.493 &0.722 &0.708 &0.645 &0.827 &0.801 &0.749 &0.890 \\
Laplace &0.694 &0.624 &0.823 &\textbf{0.770} &0.715 &0.875 &0.851 &0.816 &0.920 &0.581 &0.503 &0.729 &0.713 &0.650 &0.831 &0.803 &0.754 &0.892 \\
\bottomrule
\end{tabular*}
\label{table_different_distribution}
\end{table*}

%In this subsection, we answer \textbf{RQ2} from three aspects, including different variants of MMIEA, different of modalities and different of distribution methods. The corresponding experimental results are shown in Table~\ref{table_different_module} and Table~\ref{table_different_distribution}.
\smallskip 
\noindent \textbf{Impact of Modalities.}
%\subsubsection{Different of Modalities.} 
\vic{The upper part of Table~\ref{table_different_module} shows the individual contribution of different modalities. We can observe that independent of the  dataset or the number of pre-aligned seeds, the removal of different modalities has varying degrees of performance drop. The structural information has shown to be the main source, with its removal leading to the most significant drop (this is in line with previous findings~\citep{Zhenxi_2022}). This might be explained by the wealth of structural triples available in both datasets. On the other extreme, the performance gain brought by the visual modality is minimal. In fact, the  removal of  visual information can sometimes lead to achieve better results. The main reason is that the visual information provides limited additional knowledge. Only through the interaction with other modal information can bring  certain performance improvement.}
%The upper part of Table~\ref{table_different_module} shows the individual contribution of different modalities, we can observe that no matter which dataset is used or how large the pre-aligned seeds are, the removal of different modalities has varying degrees of performance drop, and the structure modal has shown the primary importance with the most significant drop, which is in line with the previous work~\citep{Zhenxi_2022}. This may be a consequence of that both datasets provides a wealth of structured triple knowledge. After encoding with GAT, entities can obtain more semantically rich representations, thus enabling better exploration of the correlation between entities. Recognizing the importance of structured information, we focus on inter-modal interactions with a structure modality at the core in PMF module. Furthermore, the performance gain brought by the visual modality is minimal, and even removing the visual modality can achieve better results in some cases. The main reason is that as an additional supplementary content, visual information provides relatively limited knowledge. Only through intersection with other modal information can a certain performance improvement be achieved.
%\subsubsection{Different Variants of MMIEA} 

\smallskip
\noindent \textbf{Impact of Modules.} 
\vic{The lower part of Table~\ref{table_different_module} presents the results of the impact of each component of MIMEA on the performance. We can observe  that by removing any module the performance  dramatically degrades.
%the performance when removing any module from MMIEA, which indicates the necessity of each module. 
This could be explained by the fact that different modules play different roles, realizing multi-granular modal information interaction. For example, the PMF module focuses on the interaction of uni-modal information (with the structural information as the core) and can ultimately form  joint-modal representations. 
% The OTMA module pays attention to the interaction between uni-modal and joint-modal. 
In contrast, the MCL module underscores the significance of intra-modal interactions for each modality.  The MIMEA's modules are interrelated and form a complete data flow, so the absence of any one of them leads to a significant performance fluctuation.}

%The lower part of Table~\ref{table_different_module} presents the results on the impact of each component of MMIEA. We can find that it dramatically degrades the performance when removing any module from MMIEA, which indicates the necessity of each module. This can be mainly explained that different modules play different roles, each module performs its own duties and can realize multi-granularity modal information interaction. For example, PMF module focuses more on the interaction capabilities between uni-modal with the structural modal as the core, and can ultimately form the joint-modal representations of multi-modal. OTMA module places its emphasis on the interaction between uni-modal and the joint-modal representations. In contrast, MCL module underscores the significance of intra-modal interactions within individual modalities. These three modules are interrelated and form a complete data flow, so the absence of any one of them can lead to a significant performance fluctuation.

\smallskip 
\noindent \textbf{Impact of Distribution Methods.}
%\subsubsection{Different of Distribution Methods}
\vic{We investigate the choice of different probability distribution functions in the PMF module. Table~\ref{table_different_distribution} reports the results by replacing the Beta function in the PMF module with the Cauchy, Gamma, Gumbel, or Laplace functions. We observe that using different probability distribution functions has a relatively limited impact on MIMEA's performance, showing the robustness of the PMF module. This is explained by the fact that the weight coefficients obtained by each probability distribution function tend to be  similar after subsequent gradient updates.}
%To further investigate the impact of different probability distribution functions in the PMF module, we report the results by replacing the Beta function in PMF module with Cauchy, Gamma, Gumbel, and Laplace function, as shown in Table~\ref{table_different_distribution}. We can observe that using different probability distribution functions has a relatively limited impact on model performance, and the performance differences among these different probability distribution functions are not substantial. It demonstrates the robustness of the PMF module towards the selection of probability distribution functions. The possible reason is that the probability function in the PMF module is primarily a means to obtain weight coefficient for individual modalities in the joint modality representation. The weight coefficients obtained by each probability distribution function tend to be relatively similar after subsequent gradient updates, thereby mitigating the importance of the specific choice of the probability function for obtaining these weight coefficients.

\smallskip 
\noindent {\textbf{Impact of Different Pivotal Modality.}
In the PMF module, we use the structural modality as the central one for the interaction between uni-modal representations. To verify the adequateness  of this choice, we select attribute, relation, and visual as the central ones. The experimental results are shown in Table~\ref{table_different_core}. We can observe that by choosing the structural modality as the core we achieve the best results. The main reason is that the datasets contain rich knowledge of structural triples, which can provide abundant evidence. Recall  that the performance loss caused by removing the visual modality in Table~\ref{table_different_module} is lower than that of removing the relation and attribute modalities, that is, the visual modality seems to be of little importance in the MMEA task. However, when using the visual modality as the core for uni-modal interaction, it can achieve better results than the relation and attribute modalities. A possible explanation is that the subsequent OTMA module directly assists the functioning  of the visual modality, because from Table~\ref{table_different_module} we find that removing the OTMA module has the largest  impact on performance.
}

\begin{table}[!htp]
\setlength{\abovecaptionskip}{0.05cm}
\renewcommand\arraystretch{1.3}
\setlength{\tabcolsep}{0.27em}
\centering
\small
\caption{\vic{The MRR metric results of using different modal content as the central one in the PMF module.} Best scores are highlighted in \textbf{bold}.}
\begin{tabular*}{0.72\linewidth}{@{}ccccccc@{}}
\bottomrule
\multicolumn{1}{c}{\multirow{2}{*}{\textbf{Methods}}} & \multicolumn{3}{c}{\textbf{FB15K-DB15K}} & \multicolumn{3}{c}{\textbf{FB15K-YAGO15K}}\\
\cline{2-3}\cline{4-5}\cline{6-7}
& \multicolumn{1}{c}{\textbf{20\%}} & \multicolumn{1}{c}{\textbf{50\%}} & \multicolumn{1}{c}{\textbf{80\%}} & \multicolumn{1}{c}{\textbf{20\%}} & \multicolumn{1}{c}{\textbf{50\%}} & \multicolumn{1}{c}{\textbf{80\%}} \\

\hline
attribute &0.595 &0.747 &0.841 &0.494 &0.688 &0.794 \\
relation &0.576 &0.724 &0.845 &0.518 &0.660 &0.796 \\
visual &0.630 &0.744 &0.844 &0.535 &0.680 &0.780 \\
structural &\textbf{0.694} &\textbf{0.770} &\textbf{0.855} &\textbf{0.587} &\textbf{0.712} &\textbf{0.803} \\
\bottomrule
\end{tabular*}
\label{table_different_core}
\end{table}

\subsection{Complexity Analysis}
\hypertarget{complexity_analysis}{}
\vic{
To address \textbf{RQ3}, we analyze the model's complexity from two perspectives: time complexity and space complexity. The time complexity can be measured by the amount of model calculations, while the space complexity can be measured by the amount of model's parameters. Model calculation volume refers to the number of floating-point operations performed during the inference process of the model, usually expressed in units of FLOPs (Floating-Point Operations Per Second). The number of model parameters refers to the number of adjustable parameters that need to be learned in the model. These parameters are the weights and biases of the model that are adjusted through optimization algorithms such as gradient descent during the training process. The number of parameters is usually expressed in  ``Millions'' (M) or ``Billions'' (G). Table~\ref{table_complexity_analysis} presents the time and space complexity results  of MIMEA and the best-performing MCLEA~\citep{Zhenxi_2022} and the MEAformer~\citep{Zhuo_2022} model. We can find that MIMEA simultaneously reduces the computational cost and the number of parameters in comparison to the other two baselines. In particular,  the amount of calculation needed by MIMEA is one-third of MEAformer's. To sum up,  MIMEA can achieve the best performance while minimizing the model's computational load and video memory footprint.}

\begin{table}[!htp]
\setlength{\abovecaptionskip}{0.03cm}
\renewcommand\arraystretch{1.2}
\setlength{\tabcolsep}{0.3em}
\centering
\small
\caption{Amount of parameters and calculations required by different models on different datasets.}
\begin{tabular*}{0.9\linewidth}{@{}cccc@{}}
\bottomrule
\multicolumn{1}{c}{\textbf{Metrics}}&\multicolumn{1}{c}{\textbf{Model}}&\multicolumn{1}{c}{\textbf{FB15K-DB15K}}&\multicolumn{1}{c}{\textbf{FB15K-YAGO15K}} \\
\hline
\multirow{3}{*}{FLOPs} & MCLEA~\citep{Zhenxi_2022} &103.345G &112.872G \\
& MEAformer~\citep{Zhuo_2022} &203.100G &219.175G \\
& MIMEA &67.770G &74.018G \\
\hline
\multirow{3}{*}{Params} & MCLEA~\citep{Zhenxi_2022} &3.720M &3.720M \\
& MEAformer~\citep{Zhuo_2022} &3.461M &3.374M\\
& MIMEA &2.440M &2.440M \\
\bottomrule
\end{tabular*}
\label{table_complexity_analysis}
\end{table}

\section{Conclusion and future work}
\vic{In this paper, we proposed MIMEA, a  framework for multi-modal entity alignment that effectively  leverages multi-modal knowledge with 
% \textcolor{red}{the exploitation of intra-modal and inter-modal effect} 
 the exploitation of intra-modal and inter-modal interactions. %Specifically, we first utilizes multiple individual encoders to obtain modality-specific representations for each entity. Then, we employ the probability distribution mechanism to integrate heterogeneous uni-modal knowledge into joint-modal representations, and further introduce the optimal transport strategy to facilitate interaction between uni-modal and joint-modal embeddings. Moreover, we introduce the contrastive learning manner to explore the intra-modal relationships. 
The experimental results demonstrate the effectiveness of MIMEA. For future work, 
% we will focus on incorporating additional modal information, such as textual descriptions of entities.
%, such as textual descriptions of entities, to further enhance entity alignment performance. 
% Additionally, 
given that the structural information is the most significant and  that in practice, structural knowledge is often incomplete, we could first employ knowledge graph completion techniques to fill in missing parts.}
%In this paper, we proposed MMIEA, a novel framework for multi-modal entity alignment, to comprehensively leverage multi-modal knowledge with the exploitation of intra-modal and inter-modal effect. Specifically, we first utilizes multiple individual encoders to obtain modality-specific representations for each entity. Then, we employ the probability distribution mechanism to integrate heterogeneous uni-modal knowledge into joint-modal representations, and further introduce the optimal transport strategy to facilitate interaction between uni-modal and joint-modal embeddings. Moreover, we introduce the contrastive learning manner to explore the intra-modal relationships. The extensive experiments and ablation studies results demonstrate the effectiveness of MMIEA. For future work, we will focus on incorporating additional modal information, such as textual descriptions of entities, to further enhance entity alignment performance. Additionally, given that structured modality has the most significant impact on the model's performance, and in practice, structured knowledge is often incomplete, we can first employ knowledge graph completion techniques to fill in missing parts of structured knowledge before proceeding with further encoding of the structured content.

\bibliographystyle{ACM-Reference-Format}
\bibliography{sample-base}


\begin{thebibliography}{42}

%%% the Web address.

\ifx \showCODEN    \undefined \def \showCODEN     #1{\unskip}     \fi
\ifx \showDOI      \undefined \def \showDOI       #1{#1}\fi
\ifx \showISBNx    \undefined \def \showISBNx     #1{\unskip}     \fi
\ifx \showISBNxiii \undefined \def \showISBNxiii  #1{\unskip}     \fi
\ifx \showISSN     \undefined \def \showISSN      #1{\unskip}     \fi
\ifx \showLCCN     \undefined \def \showLCCN      #1{\unskip}     \fi
\ifx \shownote     \undefined \def \shownote      #1{#1}          \fi
\ifx \showarticletitle \undefined \def \showarticletitle #1{#1}   \fi
\ifx \showURL      \undefined \def \showURL       {\relax}        \fi
% invisible to TeX
\providecommand\bibfield[2]{#2}
\providecommand\bibinfo[2]{#2}
\providecommand\natexlab[1]{#1}
\providecommand\showeprint[2][]{arXiv:#2}

\bibitem[Asano et~al\mbox{.}(2020)]%
        {Yuki_2020}
\bibfield{author}{\bibinfo{person}{Yuki~Markus Asano}, \bibinfo{person}{Mandela
  Patrick}, \bibinfo{person}{Christian Rupprecht}, {and}
  \bibinfo{person}{Andrea Vedaldi}.} \bibinfo{year}{2020}\natexlab{}.
\newblock \showarticletitle{Labelling unlabelled videos from scratch with
  multi-modal self-supervision}. In \bibinfo{booktitle}{\emph{NeurIPS}}.
  \bibinfo{publisher}{Curran Associates}, \bibinfo{address}{online},
  \bibinfo{pages}{1--12}.
\newblock


\bibitem[Bonneel et~al\mbox{.}(2011)]%
        {Nicolas_2011}
\bibfield{author}{\bibinfo{person}{Nicolas Bonneel}, \bibinfo{person}{Michiel
  van~de Panne}, \bibinfo{person}{Sylvain Paris}, {and}
  \bibinfo{person}{Wolfgang Heidrich}.} \bibinfo{year}{2011}\natexlab{}.
\newblock \showarticletitle{Displacement Interpolation using Lagrangian Mass
  Transport}.
\newblock \bibinfo{journal}{\emph{{ACM} Trans. Graph.}} \bibinfo{volume}{30},
  \bibinfo{number}{6} (\bibinfo{year}{2011}), \bibinfo{pages}{158}.
\newblock


\bibitem[Cao et~al\mbox{.}(2019)]%
        {Yixin_2019}
\bibfield{author}{\bibinfo{person}{Yixin Cao}, \bibinfo{person}{Zhiyuan Liu},
  \bibinfo{person}{Chengjiang Li}, \bibinfo{person}{Zhiyuan Liu},
  \bibinfo{person}{Juanzi Li}, {and} \bibinfo{person}{Tat{-}Seng Chua}.}
  \bibinfo{year}{2019}\natexlab{}.
\newblock \showarticletitle{Multi-Channel Graph Neural Network for Entity
  Alignment}. In \bibinfo{booktitle}{\emph{ACL}}. \bibinfo{publisher}{ACL},
  \bibinfo{address}{Florence, Italy}, \bibinfo{pages}{1452--1461}.
\newblock


\bibitem[Cao et~al\mbox{.}(2022)]%
        {Zongsheng_2022}
\bibfield{author}{\bibinfo{person}{Zongsheng Cao}, \bibinfo{person}{Qianqian
  Xu}, \bibinfo{person}{Zhiyong Yang}, \bibinfo{person}{Yuan He},
  \bibinfo{person}{Xiaochun Cao}, {and} \bibinfo{person}{Qingming Huang}.}
  \bibinfo{year}{2022}\natexlab{}.
\newblock \showarticletitle{{OTKGE:} Multi-modal Knowledge Graph Embeddings via
  Optimal Transport}. In \bibinfo{booktitle}{\emph{NeurIPS}}.
  \bibinfo{publisher}{NeurIPS Foundation}, \bibinfo{address}{New Orleans, the
  United States}, \bibinfo{pages}{1--13}.
\newblock
\urldef\tempurl%
\url{http://papers.nips.cc/paper\_files/paper/2022/hash/ffdb280e7c7b4c4af30e04daf5a84b98-Abstract-Conference.html}
\showURL{%
\tempurl}


\bibitem[Caron et~al\mbox{.}(2020)]%
        {Mathilde_2020}
\bibfield{author}{\bibinfo{person}{Mathilde Caron}, \bibinfo{person}{Ishan
  Misra}, \bibinfo{person}{Julien Mairal}, \bibinfo{person}{Priya Goyal},
  \bibinfo{person}{Piotr Bojanowski}, {and} \bibinfo{person}{Armand Joulin}.}
  \bibinfo{year}{2020}\natexlab{}.
\newblock \showarticletitle{Unsupervised Learning of Visual Features by
  Contrasting Cluster Assignments}. In \bibinfo{booktitle}{\emph{NeurIPS}}.
  \bibinfo{publisher}{Curran Associates}, \bibinfo{address}{online},
  \bibinfo{pages}{1--13}.
\newblock


\bibitem[Chang et~al\mbox{.}(2022)]%
        {Wanxing_2022}
\bibfield{author}{\bibinfo{person}{Wanxing Chang}, \bibinfo{person}{Ye Shi},
  \bibinfo{person}{Hoang Tuan}, {and} \bibinfo{person}{Jingya Wang}.}
  \bibinfo{year}{2022}\natexlab{}.
\newblock \showarticletitle{Unified Optimal Transport Framework for Universal
  Domain Adaptation}. In \bibinfo{booktitle}{\emph{NeurIPS}}.
  \bibinfo{publisher}{Curran Associates}, \bibinfo{address}{New Orleans, LA,
  USA}, \bibinfo{pages}{1--13}.
\newblock


\bibitem[Chen et~al\mbox{.}(2020)]%
        {Liyi_2020}
\bibfield{author}{\bibinfo{person}{Liyi Chen}, \bibinfo{person}{Zhi Li},
  \bibinfo{person}{Yijun Wang}, \bibinfo{person}{Tong Xu},
  \bibinfo{person}{Zhefeng Wang}, {and} \bibinfo{person}{Enhong Chen}.}
  \bibinfo{year}{2020}\natexlab{}.
\newblock \showarticletitle{{MMEA:} Entity Alignment for Multi-modal Knowledge
  Graph}. In \bibinfo{booktitle}{\emph{KSEM}}. \bibinfo{publisher}{Springer},
  \bibinfo{address}{Hangzhou, China}, \bibinfo{pages}{134--147}.
\newblock


\bibitem[Chen et~al\mbox{.}(2022b)]%
        {Liyi_2022}
\bibfield{author}{\bibinfo{person}{Liyi Chen}, \bibinfo{person}{Zhi Li},
  \bibinfo{person}{Tong Xu}, \bibinfo{person}{Han Wu}, \bibinfo{person}{Zhefeng
  Wang}, \bibinfo{person}{Nicholas~Jing Yuan}, {and} \bibinfo{person}{Enhong
  Chen}.} \bibinfo{year}{2022}\natexlab{b}.
\newblock \showarticletitle{Multi-modal Siamese Network for Entity Alignment}.
  In \bibinfo{booktitle}{\emph{KDD}}. \bibinfo{publisher}{ACM},
  \bibinfo{address}{Washington, the United States}, \bibinfo{pages}{118--126}.
\newblock


\bibitem[Chen et~al\mbox{.}(2018)]%
        {Muhao_2018}
\bibfield{author}{\bibinfo{person}{Muhao Chen}, \bibinfo{person}{Yingtao Tian},
  \bibinfo{person}{KaiWei Chang}, \bibinfo{person}{Steven Skiena}, {and}
  \bibinfo{person}{Carlo Zaniolo}.} \bibinfo{year}{2018}\natexlab{}.
\newblock \showarticletitle{Co-training Embeddings of Knowledge Graphs and
  Entity Descriptions for Cross-lingual Entity Alignment}. In
  \bibinfo{booktitle}{\emph{IJCAI}}. \bibinfo{publisher}{ijcai.org},
  \bibinfo{address}{Stockholm, Sweden}, \bibinfo{pages}{3998--4004}.
\newblock


\bibitem[Chen et~al\mbox{.}(2017)]%
        {Muhao_2017}
\bibfield{author}{\bibinfo{person}{Muhao Chen}, \bibinfo{person}{Yingtao Tian},
  \bibinfo{person}{Mohan Yang}, {and} \bibinfo{person}{Carlo Zaniolo}.}
  \bibinfo{year}{2017}\natexlab{}.
\newblock \showarticletitle{Multilingual Knowledge Graph Embeddings for
  Cross-lingual Knowledge Alignment}. In \bibinfo{booktitle}{\emph{IJCAI}}.
  \bibinfo{publisher}{ijcai.org}, \bibinfo{address}{Melbourne, Australia},
  \bibinfo{pages}{1511--1517}.
\newblock


\bibitem[Chen et~al\mbox{.}(2022a)]%
        {Zhuo_2022}
\bibfield{author}{\bibinfo{person}{Zhuo Chen}, \bibinfo{person}{Jiaoyan Chen},
  \bibinfo{person}{Wen Zhang}, \bibinfo{person}{Lingbing Guo},
  \bibinfo{person}{Yin Fang}, \bibinfo{person}{Yufeng Huang},
  \bibinfo{person}{Yuxia Geng}, \bibinfo{person}{Jeff~Z. Pan},
  \bibinfo{person}{Wenting Song}, {and} \bibinfo{person}{Huajun Chen}.}
  \bibinfo{year}{2022}\natexlab{a}.
\newblock \showarticletitle{MEAformer: Multi-modal Entity Alignment Transformer
  for Meta Modality Hybrid}.
\newblock \bibinfo{journal}{\emph{CoRR}}  \bibinfo{volume}{abs/2212.14454}
  (\bibinfo{year}{2022}), \bibinfo{pages}{1--11}.
\newblock


\bibitem[Cuturi(2013)]%
        {Marco_2013}
\bibfield{author}{\bibinfo{person}{Marco Cuturi}.}
  \bibinfo{year}{2013}\natexlab{}.
\newblock \showarticletitle{Sinkhorn Distances: Lightspeed Computation of
  Optimal Transport}. In \bibinfo{booktitle}{\emph{NeurIPS}}.
  \bibinfo{publisher}{Curran Associates}, \bibinfo{address}{Lake Tahoe, the
  United States}, \bibinfo{pages}{2292--2300}.
\newblock


\bibitem[Gu et~al\mbox{.}(2022)]%
        {Xiang_2022}
\bibfield{author}{\bibinfo{person}{Xiang Gu}, \bibinfo{person}{Yucheng Yang},
  \bibinfo{person}{Wei Zeng}, \bibinfo{person}{Jian Sun}, {and}
  \bibinfo{person}{Zongben Xu}.} \bibinfo{year}{2022}\natexlab{}.
\newblock \showarticletitle{Keypoint-Guided Optimal Transport with Applications
  in Heterogeneous Domain Adaptation}. In \bibinfo{booktitle}{\emph{NeurIPS}}.
  \bibinfo{publisher}{Curran Associates}, \bibinfo{address}{New Orleans, LA,
  USA}, \bibinfo{pages}{1--14}.
\newblock


\bibitem[Gu et~al\mbox{.}(2019)]%
        {GZCL_2019}
\bibfield{author}{\bibinfo{person}{Yu Gu}, \bibinfo{person}{Tianshuo Zhou},
  \bibinfo{person}{Gong Cheng}, \bibinfo{person}{Ziyang Li},
  \bibinfo{person}{Jeff~Z. Pan}, {and} \bibinfo{person}{Yuzhong Qu}.}
  \bibinfo{year}{2019}\natexlab{}.
\newblock \showarticletitle{{Relevance Search over Schema-Rich Knowledge
  Graphs}}. In \bibinfo{booktitle}{\emph{WSDM}}. \bibinfo{publisher}{ACM
  Press}, \bibinfo{address}{Melbourne, Australia}, \bibinfo{pages}{114--122}.
\newblock


\bibitem[Guo et~al\mbox{.}(2021)]%
        {Hao_2021}
\bibfield{author}{\bibinfo{person}{Hao Guo}, \bibinfo{person}{Jiuyang Tang},
  \bibinfo{person}{Weixin Zeng}, \bibinfo{person}{Xiang Zhao}, {and}
  \bibinfo{person}{Li Liu}.} \bibinfo{year}{2021}\natexlab{}.
\newblock \showarticletitle{Multi-modal entity alignment in hyperbolic space}.
\newblock \bibinfo{journal}{\emph{Neurocomputing}}  \bibinfo{volume}{461}
  (\bibinfo{year}{2021}), \bibinfo{pages}{598--607}.
\newblock


\bibitem[Guo et~al\mbox{.}(2023)]%
        {Lingbing_2023}
\bibfield{author}{\bibinfo{person}{Lingbing Guo}, \bibinfo{person}{Zhuo Chen},
  \bibinfo{person}{Jiaoyan Chen}, {and} \bibinfo{person}{Huajun Chen}.}
  \bibinfo{year}{2023}\natexlab{}.
\newblock \showarticletitle{Revisit and Outstrip Entity Alignment: {A}
  Perspective of Generative Models}.
\newblock \bibinfo{journal}{\emph{CoRR}}  \bibinfo{volume}{abs/2305.14651}
  (\bibinfo{year}{2023}), \bibinfo{pages}{1--18}.
\newblock


\bibitem[Guo et~al\mbox{.}(2022)]%
        {Lingbing_2022}
\bibfield{author}{\bibinfo{person}{Lingbing Guo}, \bibinfo{person}{Qiang
  Zhang}, \bibinfo{person}{Zequn Sun}, \bibinfo{person}{Mingyang Chen},
  \bibinfo{person}{Wei Hu}, {and} \bibinfo{person}{Huajun Chen}.}
  \bibinfo{year}{2022}\natexlab{}.
\newblock \showarticletitle{Understanding and Improving Knowledge Graph
  Embedding for Entity Alignment}. In \bibinfo{booktitle}{\emph{ICML}}.
  \bibinfo{publisher}{{PMLR}}, \bibinfo{address}{Maryland, the United States},
  \bibinfo{pages}{8145--8156}.
\newblock


\bibitem[Hu et~al\mbox{.}(2022)]%
        {Zhiwei_2022}
\bibfield{author}{\bibinfo{person}{Zhiwei Hu}, \bibinfo{person}{V{\'{\i}}ctor
  Guti{\'{e}}rrez{-}Basulto}, \bibinfo{person}{Zhiliang Xiang},
  \bibinfo{person}{Xiaoli Li}, \bibinfo{person}{Ru Li}, {and}
  \bibinfo{person}{Jeff~Z. Pan}.} \bibinfo{year}{2022}\natexlab{}.
\newblock \showarticletitle{Type-aware Embeddings for Multi-Hop Reasoning over
  Knowledge Graphs}. In \bibinfo{booktitle}{\emph{IJCAI}}.
  \bibinfo{publisher}{ijcai.org}, \bibinfo{address}{Vienna, Austria},
  \bibinfo{pages}{3078--3084}.
\newblock


\bibitem[Kendall et~al\mbox{.}(2018)]%
        {Alex_2018}
\bibfield{author}{\bibinfo{person}{Alex Kendall}, \bibinfo{person}{Yarin Gal},
  {and} \bibinfo{person}{Roberto Cipolla}.} \bibinfo{year}{2018}\natexlab{}.
\newblock \showarticletitle{Multi-Task Learning Using Uncertainty to Weigh
  Losses for Scene Geometry and Semantics}. In
  \bibinfo{booktitle}{\emph{CVPR}}. \bibinfo{publisher}{{IEEE} Computer
  Society}, \bibinfo{address}{Salt Lake City, the United States},
  \bibinfo{pages}{7482--7491}.
\newblock


\bibitem[Lehmann et~al\mbox{.}(2015)]%
        {Jens_2015}
\bibfield{author}{\bibinfo{person}{Jens Lehmann}, \bibinfo{person}{Robert
  Isele}, \bibinfo{person}{Max Jakob}, \bibinfo{person}{Anja Jentzsch},
  \bibinfo{person}{Dimitris Kontokostas}, \bibinfo{person}{Pablo~N. Mendes},
  \bibinfo{person}{Sebastian Hellmann}, \bibinfo{person}{Mohamed Morsey},
  \bibinfo{person}{Patrick van Kleef}, \bibinfo{person}{S{\"{o}}ren Auer},
  {and} \bibinfo{person}{Christian Bizer}.} \bibinfo{year}{2015}\natexlab{}.
\newblock \showarticletitle{DBpedia - {A} large-scale, multilingual knowledge
  base extracted from Wikipedia}.
\newblock \bibinfo{journal}{\emph{Semantic Web}} \bibinfo{volume}{6},
  \bibinfo{number}{2} (\bibinfo{year}{2015}), \bibinfo{pages}{167--195}.
\newblock


\bibitem[Li et~al\mbox{.}(2019)]%
        {Chengjiang_2019}
\bibfield{author}{\bibinfo{person}{Chengjiang Li}, \bibinfo{person}{Yixin Cao},
  \bibinfo{person}{Lei Hou}, \bibinfo{person}{Jiaxin Shi},
  \bibinfo{person}{Juanzi Li}, {and} \bibinfo{person}{Tat{-}Seng Chua}.}
  \bibinfo{year}{2019}\natexlab{}.
\newblock \showarticletitle{Semi-supervised Entity Alignment via Joint
  Knowledge Embedding Model and Cross-graph Model}. In
  \bibinfo{booktitle}{\emph{EMNLP}}. \bibinfo{publisher}{ACL},
  \bibinfo{address}{Hong Kong, China}, \bibinfo{pages}{2723--2732}.
\newblock


\bibitem[Li et~al\mbox{.}(2023)]%
        {Qian_2023}
\bibfield{author}{\bibinfo{person}{Qian Li}, \bibinfo{person}{Shu Guo},
  \bibinfo{person}{Yangyifei Luo}, \bibinfo{person}{Cheng Ji},
  \bibinfo{person}{Lihong Wang}, \bibinfo{person}{Jiawei Sheng}, {and}
  \bibinfo{person}{Jianxin Li}.} \bibinfo{year}{2023}\natexlab{}.
\newblock \showarticletitle{Attribute-Consistent Knowledge Graph Representation
  Learning for Multi-Modal Entity Alignment}. In
  \bibinfo{booktitle}{\emph{WWW}}. \bibinfo{publisher}{{ACM}},
  \bibinfo{address}{Washington, the United States},
  \bibinfo{pages}{2499--2508}.
\newblock


\bibitem[Lin et~al\mbox{.}(2022)]%
        {Zhenxi_2022}
\bibfield{author}{\bibinfo{person}{Zhenxi Lin}, \bibinfo{person}{Ziheng Zhang},
  \bibinfo{person}{Meng Wang}, \bibinfo{person}{Yinghui Shi},
  \bibinfo{person}{Xian Wu}, {and} \bibinfo{person}{Yefeng Zheng}.}
  \bibinfo{year}{2022}\natexlab{}.
\newblock \showarticletitle{Multi-modal Contrastive Representation Learning for
  Entity Alignment}. In \bibinfo{booktitle}{\emph{COLING}}.
  \bibinfo{publisher}{ACL}, \bibinfo{address}{online},
  \bibinfo{pages}{2572--2584}.
\newblock


\bibitem[Liu et~al\mbox{.}(2021)]%
        {Fangyu_2021}
\bibfield{author}{\bibinfo{person}{Fangyu Liu}, \bibinfo{person}{Muhao Chen},
  \bibinfo{person}{Dan Roth}, {and} \bibinfo{person}{Nigel Collier}.}
  \bibinfo{year}{2021}\natexlab{}.
\newblock \showarticletitle{Visual Pivoting for (Unsupervised) Entity
  Alignment}. In \bibinfo{booktitle}{\emph{AAAI}}. \bibinfo{publisher}{{AAAI}
  Press}, \bibinfo{address}{online}, \bibinfo{pages}{4257--4266}.
\newblock


\bibitem[Liu et~al\mbox{.}(2019)]%
        {Ye_2019}
\bibfield{author}{\bibinfo{person}{Ye Liu}, \bibinfo{person}{Hui Li},
  \bibinfo{person}{Alberto Garc{\'{\i}}a{-}Dur{\'{a}}n},
  \bibinfo{person}{Mathias Niepert}, \bibinfo{person}{Daniel
  O{\~{n}}oro{-}Rubio}, {and} \bibinfo{person}{David~S. Rosenblum}.}
  \bibinfo{year}{2019}\natexlab{}.
\newblock \showarticletitle{{MMKG:} Multi-modal Knowledge Graphs}. In
  \bibinfo{booktitle}{\emph{ESWC}}. \bibinfo{publisher}{Springer},
  \bibinfo{address}{Portorož, Slovenia}, \bibinfo{pages}{459--474}.
\newblock


\bibitem[Liu et~al\mbox{.}(2020)]%
        {Zhiyuan_2020}
\bibfield{author}{\bibinfo{person}{Zhiyuan Liu}, \bibinfo{person}{Yixin Cao},
  \bibinfo{person}{Liangming Pan}, \bibinfo{person}{Juanzi Li}, {and}
  \bibinfo{person}{Tat{-}Seng Chua}.} \bibinfo{year}{2020}\natexlab{}.
\newblock \showarticletitle{Exploring and Evaluating Attributes, Values, and
  Structures for Entity Alignment}. In \bibinfo{booktitle}{\emph{EMNLP}}.
  \bibinfo{publisher}{ACL}, \bibinfo{address}{Zurich, Switzerland},
  \bibinfo{pages}{6355--6364}.
\newblock


\bibitem[Mahdisoltani et~al\mbox{.}(2015)]%
        {Farzaneh_2015}
\bibfield{author}{\bibinfo{person}{Farzaneh Mahdisoltani},
  \bibinfo{person}{Joanna Biega}, {and} \bibinfo{person}{Fabian~M. Suchanek}.}
  \bibinfo{year}{2015}\natexlab{}.
\newblock \showarticletitle{{YAGO3:} {A} Knowledge Base from Multilingual
  Wikipedias}. In \bibinfo{booktitle}{\emph{CIDR}}.
  \bibinfo{publisher}{www.cidrdb.org}, \bibinfo{address}{Asilomar, the United
  States}, \bibinfo{pages}{1--12}.
\newblock


\bibitem[Nguyen et~al\mbox{.}(2023)]%
        {Chau_2023}
\bibfield{author}{\bibinfo{person}{Chau Nguyen}, \bibinfo{person}{Tim French},
  \bibinfo{person}{Wei Liu}, {and} \bibinfo{person}{Michael Stewart}.}
  \bibinfo{year}{2023}\natexlab{}.
\newblock \showarticletitle{SConE: Simplified Cone Embeddings with Symbolic
  Operators for Complex Logical Queries}. In \bibinfo{booktitle}{\emph{ACL}}.
  \bibinfo{publisher}{ACL}, \bibinfo{address}{Toronto, Canada},
  \bibinfo{pages}{11931--11946}.
\newblock


\bibitem[Nguyen et~al\mbox{.}(2019)]%
        {NguyenVNNP19}
\bibfield{author}{\bibinfo{person}{Dai~Quoc Nguyen}, \bibinfo{person}{Thanh
  Vu}, \bibinfo{person}{Tu~Dinh Nguyen}, \bibinfo{person}{Dat~Quoc Nguyen},
  {and} \bibinfo{person}{Dinh~Q. Phung}.} \bibinfo{year}{2019}\natexlab{}.
\newblock \showarticletitle{A Capsule Network-based Embedding Model for
  Knowledge Graph Completion and Search Personalization.}. In
  \bibinfo{booktitle}{\emph{NAACL}}. \bibinfo{publisher}{NAACL-HLT},
  \bibinfo{address}{Minneapolis, the United States},
  \bibinfo{pages}{2180--2189}.
\newblock


\bibitem[Simonyan and Zisserman(2015)]%
        {Karen_2014}
\bibfield{author}{\bibinfo{person}{Karen Simonyan} {and}
  \bibinfo{person}{Andrew Zisserman}.} \bibinfo{year}{2015}\natexlab{}.
\newblock \showarticletitle{Very Deep Convolutional Networks for Large-Scale
  Image Recognition}. In \bibinfo{booktitle}{\emph{ICLR}}.
  \bibinfo{publisher}{OpenReview.net}, \bibinfo{address}{Sainte-Maxime,
  France}, \bibinfo{pages}{1--14}.
\newblock


\bibitem[Solomon et~al\mbox{.}(2015)]%
        {Justin_2015}
\bibfield{author}{\bibinfo{person}{Justin Solomon}, \bibinfo{person}{Fernando
  de Goes}, \bibinfo{person}{Gabriel Peyr{\'{e}}}, \bibinfo{person}{Marco
  Cuturi}, \bibinfo{person}{Adrian Butscher}, \bibinfo{person}{Andy Nguyen},
  \bibinfo{person}{Tao Du}, {and} \bibinfo{person}{Leonidas~J. Guibas}.}
  \bibinfo{year}{2015}\natexlab{}.
\newblock \showarticletitle{Convolutional Wasserstein Distances: Efficient
  Optimal Transportation on Geometric Domains}.
\newblock \bibinfo{journal}{\emph{{ACM} Trans. Graph.}} \bibinfo{volume}{34},
  \bibinfo{number}{4} (\bibinfo{year}{2015}), \bibinfo{pages}{66:1--66:11}.
\newblock


\bibitem[Suh et~al\mbox{.}(2019)]%
        {Yumin_2019}
\bibfield{author}{\bibinfo{person}{Yumin Suh}, \bibinfo{person}{Bohyung Han},
  \bibinfo{person}{Wonsik Kim}, {and} \bibinfo{person}{Kyoung~Mu Lee}.}
  \bibinfo{year}{2019}\natexlab{}.
\newblock \showarticletitle{Stochastic Class-Based Hard Example Mining for Deep
  Metric Learning}. In \bibinfo{booktitle}{\emph{CVPR}}.
  \bibinfo{publisher}{{IEEE} Computer Society}, \bibinfo{address}{Long Beach,
  the United States}, \bibinfo{pages}{7251--7259}.
\newblock


\bibitem[Sun et~al\mbox{.}(2017)]%
        {Zequn_2017}
\bibfield{author}{\bibinfo{person}{Zequn Sun}, \bibinfo{person}{Wei Hu}, {and}
  \bibinfo{person}{Chengkai Li}.} \bibinfo{year}{2017}\natexlab{}.
\newblock \showarticletitle{Cross-Lingual Entity Alignment via Joint
  Attribute-Preserving Embedding}. In \bibinfo{booktitle}{\emph{ISWC}},
  Vol.~\bibinfo{volume}{10587}. \bibinfo{publisher}{Springer},
  \bibinfo{address}{Vienna, Austria}, \bibinfo{pages}{628--644}.
\newblock


\bibitem[Sun et~al\mbox{.}(2018)]%
        {Zequn_2018}
\bibfield{author}{\bibinfo{person}{Zequn Sun}, \bibinfo{person}{Wei Hu},
  \bibinfo{person}{Qingheng Zhang}, {and} \bibinfo{person}{Yuzhong Qu}.}
  \bibinfo{year}{2018}\natexlab{}.
\newblock \showarticletitle{Bootstrapping Entity Alignment with Knowledge Graph
  Embedding}. In \bibinfo{booktitle}{\emph{IJCAI}}.
  \bibinfo{publisher}{ijcai.org}, \bibinfo{address}{Stockholm, Sweden},
  \bibinfo{pages}{4396--4402}.
\newblock


\bibitem[Sun et~al\mbox{.}(2020)]%
        {Zequn_2020}
\bibfield{author}{\bibinfo{person}{Zequn Sun}, \bibinfo{person}{Chengming
  Wang}, \bibinfo{person}{Wei Hu}, \bibinfo{person}{Muhao Chen},
  \bibinfo{person}{Jian Dai}, \bibinfo{person}{Wei Zhang}, {and}
  \bibinfo{person}{Yuzhong Qu}.} \bibinfo{year}{2020}\natexlab{}.
\newblock \showarticletitle{Knowledge Graph Alignment Network with Gated
  Multi-Hop Neighborhood Aggregation}. In \bibinfo{booktitle}{\emph{AAAI}}.
  \bibinfo{publisher}{{AAAI} Press}, \bibinfo{address}{California, the United
  States}, \bibinfo{pages}{222--229}.
\newblock


\bibitem[Velickovic et~al\mbox{.}(2018)]%
        {Petar_2018}
\bibfield{author}{\bibinfo{person}{Petar Velickovic}, \bibinfo{person}{Guillem
  Cucurull}, \bibinfo{person}{Arantxa Casanova}, \bibinfo{person}{Adriana
  Romero}, \bibinfo{person}{Pietro Li{\`{o}}}, {and} \bibinfo{person}{Yoshua
  Bengio}.} \bibinfo{year}{2018}\natexlab{}.
\newblock \showarticletitle{Graph Attention Networks}. In
  \bibinfo{booktitle}{\emph{ICLR}}. \bibinfo{publisher}{OpenReview.net},
  \bibinfo{address}{Vancouver, Canada}, \bibinfo{pages}{1--12}.
\newblock


\bibitem[Xu and Chen(2023)]%
        {Yingxue_2023}
\bibfield{author}{\bibinfo{person}{Yingxue Xu} {and} \bibinfo{person}{Hao
  Chen}.} \bibinfo{year}{2023}\natexlab{}.
\newblock \showarticletitle{Multimodal Optimal Transport-based Co-Attention
  Transformer with Global Structure Consistency for Survival Prediction}. In
  \bibinfo{booktitle}{\emph{ICCV}}. \bibinfo{publisher}{{IEEE}},
  \bibinfo{address}{Paris, France}, \bibinfo{pages}{21184--21194}.
\newblock


\bibitem[Zhang et~al\mbox{.}(2021)]%
        {Zhanqiu_2021}
\bibfield{author}{\bibinfo{person}{Zhanqiu Zhang}, \bibinfo{person}{Jie Wang},
  \bibinfo{person}{Jiajun Chen}, \bibinfo{person}{Shuiwang Ji}, {and}
  \bibinfo{person}{Feng Wu}.} \bibinfo{year}{2021}\natexlab{}.
\newblock \showarticletitle{ConE: Cone Embeddings for Multi-Hop Reasoning over
  Knowledge Graphs}. In \bibinfo{booktitle}{\emph{NeurIPS}}.
  \bibinfo{publisher}{Curran Associates}, \bibinfo{address}{online},
  \bibinfo{pages}{19172--19183}.
\newblock


\bibitem[Zhu et~al\mbox{.}(2017)]%
        {Hao_2017}
\bibfield{author}{\bibinfo{person}{Hao Zhu}, \bibinfo{person}{Ruobing Xie},
  \bibinfo{person}{Zhiyuan Liu}, {and} \bibinfo{person}{Maosong Sun}.}
  \bibinfo{year}{2017}\natexlab{}.
\newblock \showarticletitle{Iterative Entity Alignment via Joint Knowledge
  Embeddings}. In \bibinfo{booktitle}{\emph{IJCAI}}.
  \bibinfo{publisher}{ijcai.org}, \bibinfo{address}{Melbourne, Australia},
  \bibinfo{pages}{4258--4264}.
\newblock


\bibitem[Zhu et~al\mbox{.}(2021a)]%
        {Yao_2021}
\bibfield{author}{\bibinfo{person}{Yao Zhu}, \bibinfo{person}{Hongzhi Liu},
  \bibinfo{person}{Zhonghai Wu}, {and} \bibinfo{person}{Yingpeng Du}.}
  \bibinfo{year}{2021}\natexlab{a}.
\newblock \showarticletitle{Relation-Aware Neighborhood Matching Model for
  Entity Alignment}. In \bibinfo{booktitle}{\emph{AAAI}}.
  \bibinfo{publisher}{{AAAI} Press}, \bibinfo{address}{online},
  \bibinfo{pages}{4749--4756}.
\newblock


\bibitem[Zhu et~al\mbox{.}(2021b)]%
        {Yanqiao_2021}
\bibfield{author}{\bibinfo{person}{Yanqiao Zhu}, \bibinfo{person}{Yichen Xu},
  \bibinfo{person}{Feng Yu}, \bibinfo{person}{Qiang Liu}, \bibinfo{person}{Shu
  Wu}, {and} \bibinfo{person}{Liang Wang}.} \bibinfo{year}{2021}\natexlab{b}.
\newblock \showarticletitle{Graph Contrastive Learning with Adaptive
  Augmentation}. In \bibinfo{booktitle}{\emph{WWW}}.
  \bibinfo{publisher}{{ACM}}, \bibinfo{address}{Lisbon, Portugal},
  \bibinfo{pages}{2069--2080}.
\newblock


\bibitem[Zolfaghari et~al\mbox{.}(2021)]%
        {Mohammadreza_2021}
\bibfield{author}{\bibinfo{person}{Mohammadreza Zolfaghari},
  \bibinfo{person}{Yi Zhu}, \bibinfo{person}{Peter~V. Gehler}, {and}
  \bibinfo{person}{Thomas Brox}.} \bibinfo{year}{2021}\natexlab{}.
\newblock \showarticletitle{CrossCLR: Cross-modal Contrastive Learning For
  Multi-modal Video Representations}. In \bibinfo{booktitle}{\emph{ICCV}}.
  \bibinfo{publisher}{{IEEE}}, \bibinfo{address}{online},
  \bibinfo{pages}{1430--1439}.
\newblock


\end{thebibliography}

%\clearpage
%\subfile{sections/appendix}

\end{document}